**Limited Diffusion of Scientific Knowledge Forecasts Collapse**

Donghyun Kang[1,2]
Robert S. Danziger[3,4,5]
Jalees Rehman[3,6]
James A. Evans[1,2,7*]

[1]Department of Sociology, University of Chicago, Chicago, IL, USA.
[2]Knowledge Lab, University of Chicago, Chicago, IL, USA.
[3]Department of Medicine, University of Illinois at Chicago, Chicago, IL, USA.
[4]Department of Pharmacology, University of Illinois at Chicago, Chicago, IL, USA.
[5]Department of Physiology and Biophysics, University of Illinois at Chicago, Chicago, IL, USA.
[6]Department of Biochemistry and Molecular Genetics, University of Illinois, College of Medicine, Chicago, IL, USA.
[7]Santa Fe Institute, Santa Fe, NM, USA.

*Corresponding author. Email: jevans@uchicago.edu

**Abstract:** Market bubbles emerge when asset prices are driven unsustainably higher than asset values and shifts in belief burst them. We demonstrate the same phenomenon for biomedical knowledge when promising research receives inflated attention. We predict deflationary events by developing a diffusion index that captures whether research areas have been amplified within social and scientific bubbles or have diffused and become evaluated more broadly. We illustrate our diffusion approach contrasting the trajectories of cardiac stem cell research and cancer immunotherapy. We then trace the diffusion of unique 28,504 subfields in biomedicine comprising nearly 1.9M papers and more than 80M citations and demonstrate that limited diffusion of biomedical knowledge anticipates abrupt decreases in popularity. Our analysis emphasizes that restricted diffusion, implying a socio-epistemic bubble, leads to dramatic collapses in relevance and attention accorded to scientific knowledge.

Market bubbles emerge when widespread opinions about an asset, such as housing or securities, create self-reinforcing information that drives its price much higher than its value to society[1]. These bubbles are characterized by a swift surge in popularity, fueled by beliefs that the value may continue to rise and persist, leading to speculation. Such bubbles burst when shifts in opinion, often catalyzed by new data or events, precipitate radical discounts in pricing[2]. Science observers and researchers themselves have drawn parallels in science[3–5], which involves considerable investment in capital, attention, and other resources based on highly uncertain knowledge about the outcomes of research. This exposes science to the risk of forming bubbles analogous to financial markets[4,5]. Here, we operationalize the concept of scientific bubbles and their collapse, proposing a measurement framework and demonstrating that ideas and findings in science can experience abrupt booms and busts of popularity and credibility that may yield adverse consequences for science and scientists alike.

In the system of biomedical knowledge, citation counts have come to function as an operational currency[6,7], serving as a measure of the importance and impact of scientific work. This is also reflected by increasing interest in the development of indicators tracing emergent, disruptive, or breakthrough science and technology[8–12], which typically incorporate citation counts as key components. The citation metric manifests some distortion, however, from the inflation of citation counts with historical growth in articles[13] and the unequal size of fields[14]. Inspired by the analogy between financial and scientific bubbles, here we forecast substantial and dramatic declines in the popularity of research ideas—the bursting epistemic bubbles—as the degree to which those ideas remain concentrated within the same collection of authors, institutions, and biomedical subfields, failing to diffuse across social and scientific space despite initial popularity. We argue that this limited diffusion may indicate inflated attention to particular ideas that may not generalize or withstand broader scrutiny, ultimately leading to disappointment and disillusionment within the scientific community.

Consider the extreme but illuminating case of cardiac regeneration in biomedicine. Dr. Piero Anversa and collaborators led research in cardiac regeneration at the turn of the 21st Century by asserting the possibility of damaged heart muscle tissue after myocardial infarction with stem cells and progenitor cells drawn from the bone marrow or within the heart[15]. During Anversa and collaborators' peak productivity, they also exercised significant influence over the research narrative, sitting on editorial boards of high-profile American Heart Association journals like *Circulation Research* (Dr. Anversa alone reviewed hundreds of papers for *Circulation Research*, more than any other researcher in this period), serving on the NIH National Institute on Aging's Board of Scientific Counselors (2008-2013) and an interlocking matrix of NIH grant review panels. Nevertheless, findings from early cardiac regeneration work not only failed to generalize, but the experiments could not be replicated by other researchers[16]. This resulted in a dramatic breach of trust, the retraction of more than 30 related papers from leading journals, a marked discount in citations to the subfield, diminished confidence in the near-term prospects of cardiac regeneration, and Anversa's forced departure from Harvard. This, in turn, adversely impacted even those researchers who had been studying cardiac regeneration using more rigorous scientific approaches who had identified reproducible mechanisms underlying the phenomenon[17].

Our approach, however, aims to generalize beyond the severe research misconduct of an individual or a team of scientists. Accurate and honestly reported medical findings can still fail to generalize beyond the specific context of their initial investigation, despite optimism and hype regarding their transformative potential for medicine. More critically, as highlighted by science commentators[18] and biomedical researchers[19,20], unintended collective failures can also occur, as exemplified by the widespread use of misidentified or contaminated cell lines contributing to unjustified hype and misdirected attention and resources in the field. This phenomenon suggests the need for a more refined and multi-faceted framework to better model and evaluate the trajectories of scientific attention.

In this study, we demonstrate that fragile and overhyped biomedical findings could have been anticipated by analyzing their diffusion through the system of science. Utilizing PubMed Knowledge Graph[21], a large-scale bibliographical database, we provide a framework that considers distances between publications and their citing papers within the "scientific space" constituted by co-investigated biomedical entities and the "social space" constituted by collaborating scientists. Specifically, we develop a diffusion index to capture whether ideas have been amplified within social and scientific bubbles[22], or diffused more widely and tested for robustness across diverse research communities[23]. This approach allows us to gain insight into the diffusion of research ideas and their impact, ultimately helping us more rapidly assess the value and potential of scientific findings.

Our work demonstrates how a lack of diffusion measured by this framework—indicative of the existence of a scientific bubble—can anticipate a rapid decline in popularity as confidence bubbles burst. Applying the conceptual and measurement tools detailed below (Methods), we first compare two distinct trajectories from cardiac stem cell and cancer immunotherapy research papers. The upper panels of Fig. 1 illustrate our approach with two contrasting papers. Figs. 1a and 1b depict the diffusion and citation trajectories of an early paper[24] from Dr. Anversa's group on cardiac muscle regeneration using bone-marrow-derived cells and a seminal paper on cancer immunotherapy conducted by Dr. Honzo[25] within scientific and social spaces, respectively. Fig.

1a suggests that while cardiac stem cell research like this paper gained massive early attention, this did not sustain, manifesting fragile, overhyped ideas that could not withstand broader scrutiny across the scientific community or application across science. This is contrasted in Fig. 1b with the case of cancer immunotherapy, where research gradually diffused to distant research groups and topics before garnering significant attention.

Beyond papers, we trace the diffusion trajectories of 28,504 unique subfields in biomedicine[26], encompassing nearly 1.9 million papers and more than 80 million citations. Our analysis reveals that limited diffusion of biomedical knowledge is systematically associated with an early rise and abrupt drop in popularity. The bottom panels of Fig. 1 display the average trajectories of subfields by distinguishing those that experienced a sharp decline or collapse in scientific attention from those that did not by the end of 2019 (Methods). Furthermore, our post-hoc analyses show that the likelihood of collapses of subfields is positively associated with the concentration of publications from superstar biomedical researchers, echoing aspects of the Dr. Anversa case.

In this way, our work highlights that restricted diffusion in science can effectively capture socio-epistemic bubbles. Complementing citation dynamics with diffusion patterns enriches our identification of robust insight in biomedical science, which can be readily improved by discounting bubbles and promoting convergent results sourced through social and topical diversity.

## Results

### *Contrasting Trajectories of Cardiac Stem Cell Research and Cancer Immunotherapy*

Applying neural embedding models to MEDLINE data enables us to project all biomedical research articles onto scientific and social manifolds. As detailed in Methods and Supplementary Information (S2), this allows us to locate their relative positions within collaborative networks of scientists and biomedical entities through research. The cosine or angular distances between citing and cited research measured over social and scientific spaces aggregate into straightforward, continuous metrics of diffusion. To demonstrate the effectiveness of our approach utilizing these scientific and social spaces, we examine trajectories of two highly cited publications at the individual paper level, each drawn from Cardiac Stem Cell and Cancer Immunotherapy research, respectively.

Our first case is a research article published (PMID: 11777997) in *the New England Journal of Medicine* in 2002[24]. Led by Dr. Piero Anversa, this research supported the existence of substantial numbers of endogenous myocardial stem and progenitor cells, proposing their potential to regenerate heart muscle. This line of research initially received outsized attention because it suggested new possibilities for heart regeneration after severe myocardial infarctions involving massive tissue loss. This claim was later called into question by several researchers outside the Anversa network, however, eventually leading to the retraction of more than 30 papers by 2018 from claims of data fabrication and scientific malpractice[27].

Conversely, the second example, an article (PMID: 11015443) published in *the Journal of Experimental Medicine* in 2000[25] represents a study by a team of pioneering researchers in the field of cancer immunotherapy. Their work focuses on the inhibition of negative immune regulation and its implications for cancer treatment. The publication and subsequent work spurred the development of a broad spectrum of cancer immunology and immunotherapy research initiatives across many research groups and countries globally, laying the groundwork for what has become one of the most impactful innovations in cancer treatment.

The upper panels (**a** and **b**) of Fig. 1 visualize the contrasting temporal trajectories of these two publications in size of attention and diffusion within the scientific and social space, respectively, with 3D kernel density estimation (See Extended Data Fig.1 for 2D heatmaps). Annual diffusion indices are computed using citation data from the given and previous year. This rolling two-year window averages cosine distances between the focal and forward citing papers across social and scientific space, providing a dynamic measure of diffusion over time.

Dr. Anversa's publication experienced a meteoric rise in total citations during the first five years following debut. However, our measures indicate limited diffusion across the scientific space of distinct subfields and the social space of author teams citing the paper, which preceded a sharp decline in attention toward the paper, resembling the burst of market bubbles. In contrast, the article on Cancer Immunotherapy, which demonstrated the potential to inhibit negative immune regulation in treating cancer, gained early attention at a much slower pace. Nevertheless, the ideas ultimately diffused much more broadly, becoming one of the most influential innovations in recent cancer treatment and research. This culminated in awarding the 2018 Physiology and Medicine Nobel Prize to Drs. Tasuku Honjo and James P. Allison for advancing the scientific understanding of Cancer Immunotherapy. These contrasting cases demonstrate how our diffusion metric accounting for epistemic bubbles offers a more nuanced understanding of scientific influence than traditional citation counts, capturing the complex dynamics of diffusion through social and scientific spaces and its potential consequences.

*Knowledge Concentration Anticipates Collapse*

We elevate our analysis to the level of scientific subfields to systematically test the generalizability of our approach. We apply our framework to 28,504 unique biomedical subfields curated by Azoulay et al.[26]. Each subfield encompasses a compactly defined set of biomedical research articles using the PubMed Related Article (PMRA) algorithm[28] applied to a given seed article. This algorithm underpins the official PubMed interface, serving as a pivotal tool for researchers to locate articles related to a focal research paper, which has been fruitfully used in various studies, such as repercussions of scientific scandal on careers[29], shifts in research focus among scientists responding to NIH funding changes[30], and the negative impact of winning prizes for recipient competitors[31]. The subfields this approach allows us to identify enable us to analyze diffusion dynamics, epistemic bubbles, and collapses of scientific attention beyond selective, high-profile papers. Specifically, if work from a focal subfield is predominantly cited by research in close social and scientific proximity, the subfield's insights may not diffuse despite its seeming popularity and could retain inflated value due to local reinforcement. In other words, we anticipate that substantial and dramatic declines in the popularity of research ideas, conceptualized as knowledge 'bubbles bursting,' can be predicted by the degree to which these ideas, despite their apparent popularity, have failed to diffuse across the social and scientific space via citations.

Our primary outcome of interest is 'bubble bursting' or collapse, defined as an abrupt decline in the relevance of a given subfield for science. We time a bubble burst by comparing the standardized citation difference that a subfield garners in a given year to its performance two years prior, marking if it falls below an extreme threshold. This approach allows us to distinguish subfields that experienced deflationary bursts from those that did not by basing each standardized citation count difference against the values derived from 28,504 unique subfields (see Methods for details). We use the bottom 0.5% of the distribution of standardized citation differences as our threshold, which captures 4,480 out of 28,504 unique subfields as experiencing a collapse. To ensure the robustness of our results, we also apply thresholds of 0.25% and 0.1%, identifying 2,297 and 918 collapsed subfields respectively, and report the results from parallel analyses using these thresholds throughout the following analyses and in the Supplementary Information. Across these operationalizations, the subfields that experience a collapse also experienced a significant positive deviation from expected citation rates preceding collapse (Supplementary Information, S4.4). Fields that experience a disproportionate deflation experienced a previous inflation. In short, bubbles burst.

We compute our knowledge diffusion indices, our main predictors, for each subfield across scientific and social spaces. We identify papers published that reference at least one article within each subfield. We then calculate the average cosine distances between the referenced articles in each subfield and the citing papers with 2y rolling windows, separately for scientific and social spaces to measure scientific and social diffusion (Methods).

Using a nonparametric Cox survival model to predict the probability of bubble bursting, our estimation reveals the knowledge diffusion index as a strong leading signal preceding a sudden collapse in attention. We employ a one-year lag for our diffusion measures when associating them with the outcome of interest, collapse of attention. By splitting our observations into three groups with diffusion percentiles ranked by calendar year and subfield age—the bottom 10th percentile, the top 10th percentile, and the middle between them—Fig. 2 visualizes that diffusion in the social space forecasts the bursting of attention bubbles captured by the 0.5% threshold. The result indicates that low diffusion rates may signal poor long-term subfield survival. Conversely, high diffusion is related to subfield survival in the long term, avoiding extreme subfield-level deflationary events.

We confirm this pattern, presented in Fig. 1 and Fig. 2, with discrete-time event history models that allow us to consider temporal covariates, including field size and growth rate, total cumulative citations, citation concentration across papers, paper retractions, and unexpected deaths of elite scientists (see Methods). Our analysis consistently shows that the lower a paper's diffusion of influence, the greater the hazard that the subfield will experience an abrupt collapse of attention (Table 1 and Table S1.1). For example, as diffusion in social space reduces from one standard deviation above to one below the mean, it translates into a 74.02% (95% CI: 43.61%-110.85%) increase in the odds of experiencing a major reduction in scientific attention, accounting for subfield age, calendar year, and other covariates. Tables S1.2 and S1.3 in Supplementary Information show the estimations based on 0.25% and 0.1% thresholds to identify burst subfields.

Overall, we observe a more pronounced impact of limited social diffusion than scientific diffusion on the likelihood of subfield collapse. We posit that this likely stems from tacit confounders in research that emerge when conducted by a concentrated, connected group of scientists. When close-knit groups perform research under uniform assumptions, methodologies, and even shared resources, their findings are less likely to replicate among outsiders[23,32]. By contrast, the applicability of verified scientific findings across different biomedical domains may vary. A therapy's effectiveness for treating breast cancer is undiminished by its irrelevance for heart disease. But the failure of findings to diffuse across different groups of scientists in the same area indicates a limitation of their published scientific knowledge.

We conduct a series of subsequent analyses to gain deeper insight into characteristics of socio-epistemic "bubbles" and consequences of their collapse. Our analysis revealed that the importance of superstar scientists within a subfield, as quantified by the proportion of their publications per subfield, is positively correlated with the likelihood of collapse compared to subfields that did not burst. "Star scientists" whose work dominates a subfield[26] are more likely to have their early findings overhyped and subsequently "burst", than subfields less influenced by those dominating stars (Supplementary Information S4.1). When star scientists' findings do not diffuse corresponding with their seeming popularity, this indicates that others have attempted to generalize their results and failed. By contrast, findings from scientists with smaller reputations are more likely ignored and less likely to generate a bubble of misguided enthusiasm. Correspondingly, we find a positive association between the fraction of NIH funding allocated to collaborators of star scientists and the likelihood of attentional collapse (Supplementary Information S4.2). This suggests that limited diffusion and subsequent collapses may be correlated with the concentration of "scientific capital" in terms of reputation and resources[33], as exemplified by the Stem Cell Cardiac case discussed above. We also examine the relationship between epistemic bubbles and the limits of clinical translation. These complementary analyses demonstrate how bubbles that subsequently burst are less likely translated into clinical applications (Supplementary Information S4.3).

To evaluate the implications of epistemic bubbles and bursts, we compare the productivity of authors who published their articles close to the time of collapse (e.g., authors who published in 2001, 2002, or 2003 when the collapse was measured in 2003) with those who published in the same subfield at an earlier time (e.g., in or before 2000). As shown in Fig. 3 and Extended Data Table 1, findings suggest that those who entered right before collapse were significantly less productive in the mean number of publications both 5 and 10 years after collapse, compared to early entrants. This suggests that subfield collapse may shape researchers' reputations and career outcomes.

We also consider the implications of bubbles for the allocation of research funding. We trace the average number of new grants acknowledged per year in papers across subfields. Our analysis shows that more than 80% of the subfields, which experienced a substantial decrease in scientific attention, acknowledged new grants after collapse. By the end of 2019, the median number of such grants was 6, as detailed in Extended Data Table 2. Figure 4 illustrates the trends from 15 years before to 10 years after the collapse. It shows that while peaks of new funding precede collapse, the rate at which funding decreases after a burst is markedly slower than the trend observed before it. This pattern suggests a substantial lag by which money continues to support research that the broader biomedical community may perceive as less scientifically and clinically relevant.

**Discussion**

Current metrics of scientific attention and confidence pay scant attention to patterns of research consumption and diffusion across diverse people, institutions, disciplines, regions, and beyond. This lack of consideration can lead to an incomplete understanding of a research field's true impact and potential. Our knowledge diffusion index contrasts with and complements citation counts, the conventional unit of scientific credit. Citations alone are blind to who, where, and how far across the landscape of science those building on research reside, but our diffusion index provides a more comprehensive view.

A constriction in diffusion identifies an epistemic bubble or echo chamber that represents a leading indicator of future collapse in relevance and attention accorded to scientific and biomedical knowledge. Researchers can anticipate the collapse of biomedical approaches years prior to their occurrence by systematically tracking the diffusion of their ideas across scientists and biomedical areas. Additionally, science and biomedical policy that analyzes knowledge diffusion patterns can anticipate such collapses and may reduce their occurrence by incentivizing and accounting for diverse, disconnected support for robust scientific and medical claims[23].

Like other methods aimed at quantitatively evaluating research impact, our framework for measuring diffusion and its implementation should not replace the holistic judgment of research quality. Furthermore, while we draw on the concept of 'bubbles' in science, analogous to those in financial markets, it is worthwhile to recognize their unique aspects in the context of science. For example, small, dense research networks may be crucial for initiating high-risk projects at early stages despite a high probability of failure. In addition, scientific bubbles may not always arise from speculation, but could result from authentic scientific enthusiasm or localized beliefs in a promising research direction.

Nevertheless, our finding holds strong implications for biomedical researchers, science-based industries, and science policymakers. By accounting for diffusion and diversity, funding agencies can spot bubbles and adjust resource allocation by diversifying groups of researchers sponsored for a particular research topic. Research information platforms like PubMed, OpenAlex, the Web of Science, or Google Scholar could also incorporate strong, leading signals from which analysts can anticipate the future relevance of current research. A high diffusion index indicates that trending insights are more likely robust than fragile. Regular self-assessments of knowledge diffusion could enable individual researchers, teams, and labs to better gauge the robustness and future impact of their work. Further, documenting associations between scientific knowledge diffusion and its applications, as in the translation of biomedical research from bench to clinic, can better inform science policy.

Our results draw on subfields identified in academic science using a particular delineation of research subfields. Nevertheless, our analysis demonstrates robust evidence for the wisdom of diverse crowds in science and technology to sustain advance. It underscores the importance of both social and scientific diversity for robust evaluation of an idea's relevance to science as a whole. Moreover, our proposed framework for measuring diffusion may extend to other domains of knowledge, such as the spread of misinformation, by allowing us to measure diversity in information consumption[34]. In social media, algorithmic metrics that account for diversity in diffusion would be far less susceptible to strategic, concentrated efforts seeking to misclassify information as a legitimate, widespread trend (e.g., on Facebook's Newsfeed), just as they would decrease the intentional or unintentional illusion of scientific support.

In this way, we demonstrate the importance of idea diffusion for advancing scientific knowledge, its ability to transfer across broad science communities, and the relevance of these signals for forecasting robust ideas upon which to build novel and critical scientific and biomedical knowledge. Ultimately, our analysis underscores the relative importance of identifying the path of an idea's consumption over its point of production for predicting lasting, far-reaching impact. Accounting for this will enable the design of wise and diverse research, development, and clinical crowds, leading to improved research policy, greater reproducibility, and more sustained impact on future knowledge.

## Methods

*Manifold Representations of Social and Scientific Space*

To assess the diffusion of ideas in science from biomedicine, we train two high-dimensional vector representations using neural embedding models[35] for publications cataloged in the PubMed Knowledge Graph (PKG)[21]. The PKG provides 15,530,165 disambiguated author IDs and 481,497 unique combinations of Medical Subject Headings (MeSH) from 29,339 MeSH descriptors and 76 qualifiers, each assigned to 28,329,992 and 26,666,615 MEDLINE-indexed publications, respectively, by the end of 2019. Each document in the PubMed database is assigned a unique document identifier, PMID. The database also contains the publications to the publication reference records, which integrates PubMed's citation data, NIH's open citation collection, OpenCitations, and the Web of Science.

We specifically adapt the Doc2vec model[35], a variant of the Word2vec model[36], originally developed to produce dense vector representations for documents or paragraphs from the words that compose them. This approach has previously been extended to generate high-dimensional representational vectors geometrically proximate to the degree that entities frequently share neighbors, contexts[36–38], or are connected via social ties[39,40].

We consider that a biomedical research article can be characterized by a list of: 1) MeSH terms and 2) research collaborators. Consequently, we build two separate representational vector spaces to capture "scientific space" and "social space", respectively. For training our vector representations, we utilize the Python Gensim package[41]. We specifically use the Distributed Bag of Words (DBOW) model, analogous to the skip-gram model from the Word2vec framework, and simultaneously train the vector position of constituting elements (MeSH terms or author IDs) along with document vectors. This results in two spaces trained on 100-dimensional vector representations for PMIDs and their constituent elements. Training and validation procedures are detailed in the Supplementary Information (S2).

*Delineating Biomedical Subfields*

Biomedical knowledge obtains influence when others recognize and build on it[33,42]. In this work, we seek to understand the dynamics of diffusion and shifting attention at the level of biomedical subfields, which we define as a group of biomedical publications tightly related to a medically and biologically relevant research topic, identified through the PubMed Related Algorithm (PMRA)[28]. This method has been previously employed in studies examining the impact of publication retraction[43], repercussions of scientific scandal on careers[29], shifts in research focus by scientists in response to NIH funding changes[30], negative impacts from prize-winning on recipient competitors[31], and consequences of the premature death of elite life scientists[26] on subfields.

We specifically use the 28,504 unique seed articles curated by Azoulay et al.[26], derived from publications by "superstar"[6,44] biomedical scientists. Applying the PMRA-powered similar article function in PubMed enables us to capture over 1.9 million unique articles associated with these subfields published through 2019. We then extract ~86.8 million paper-to-paper citations identified by PKG based on them. A more comprehensive illustration of the original data source and our extension is available in Supplementary Information (S3.1). To ensure robustness, we perform complementary analyses that redefine subfields based on the position of papers within our scientific embedding space, resulting in the same pattern of findings. Details and results are reported in Supplementary Information (S3.2 and S3.3).

*Model*

Using a nonparametric Cox model and discrete-time event history model, we relate the annual diffusion indices for each subfield calculated across social and scientific spaces with an abrupt decline in the relevance of a given subfield, or "bubble burst," as illustrated in Fig. 1A. Formally, the discrete-time event history analysis model can be written as:

$$log(\frac{p_{ti}}{1-p_{ti}}) = \alpha D_{ti} + \beta x_{(t-1)i} \quad \cdots \text{Eq. (1).}$$

$p_{ti}$ denotes the probability of event happens at $t$ for subfield $i$,
$D_{ti}$ denotes time dummies corresponding to $t$ with coefficients $\alpha$,
$x_{ti}$ is vector for covariates (time varying and constant over time) with coefficients $\beta$.

***Outcome Event: Bubble bursting***

Our primary outcome of interest is the event of socio-epistemic bubbles bursting, characterized by an abrupt decline in popularity of a given subfield that we measure in the decline of citation counts as illustrated in Fig. 1. Specifically, we time bubble bursts based on when the standardized citation count difference of a given year from a subfield falls below extreme cutoffs within the life cycle of each subfield. This requires distinguishing subfields that experienced deflationary bursting, or collapse, from those that did not. We achieve this through the following steps.

We first compute $\Delta_{i,t} = c_i(t) - c_i(t-2)$, where $c_i(t)$ is the citations that a subfield $i$ garnered during year $t$ across 1970 to 2019. Unlike Azoulay et al.[26] that uses publications indexed both in Web of Science and MEDLINE, we use all PMID to PMID citation links identified in PKG 2020 data to compute citation counts. (We include all MEDLINE indexed publications, even when MeSH terms or author disambiguated IDs are not assigned to them.) Then, we standardize $\Delta_{i,t}$ within the life cycle of each subfield to make the $\Delta_{i,t}$ values comparable across 28,504 subfields. This is achieved by transforming $\Delta_{i,t}$ to $z_{i,t}$ by subtracting the mean of , $\overline{\Delta_i} = \frac{1}{N} \sum \Delta_{i,t}$, from and dividing it by the standard deviation of computed within a subfield. By doing so, we obtain the distribution of the standardized two-year citation difference, $z_{i,t}$, across 28,504 subfields. The distribution of , with the range of [-5.2, 5.52], is presented in Extended Data Fig. 2.

We operationalize bubble bursts as when the standardized citation count difference for a given year in a subfield, $z_{i,t}$, falls below extreme cutoffs, such as 0.5%, 0.25%, or 0.1% of the distribution. To qualify a decline as a burst, we require that the average of $z_{i,t}$ after the drop must be negative, ensuring a continued loss of attention. Additionally, the peak citation count at the subfield level should not occur in 2019, the final year of our dataset. If a subfield experiences more than one sharp decline, we consider the year with the most substantial one as the time of the burst. We note that bursts are preceded by bubbles: fields that experience these extreme drops also manifest greater than expected citations prior to collapse (Supplementary Information S4.4).

Using the 0.5% cutoff (i.e., ) identifies 4,480 subfields (15.7% of 28,504 subfields) that experienced a sharp decline in collective scientific attention relative to other subfields. Applying the 0.25% ( ) and 0.1% ( ) cutoffs return 2,297 and 918 subfields with the bubble bursting events, respectively. Extended Data Fig. 3 contrasts three examples of subfields that did not experience these bubbles and bursts (top panels) with three examples that exhibited substantial declines in attention (bottom panels), according to our procedure described above.

***Key Indicator: Knowledge Diffusion***

The key leading indicator for our analysis is subfield-level knowledge diffusion. We measure the knowledge diffusion by employing a 2-year rolling window approach. For each year, we identify papers published either in that year or the preceding year referencing at least one article published within a given subfield. We then separately calculate the average cosine distances (or 1-cosine similarity) between these focal articles in the subfield and the citing papers in our scientific and social spaces. This consideration leads us to measure two diffusion indices: 1) Diffusion across *Scientific Space* and 2) Diffusion across *Social Space*. In our model, we incorporate a one-year lag to assess the association between diffusion dynamics and the subsequent decline in citations.

***Further Characterization of Subfield Dynamics***

A. *Time Effect*

- ***Subfield Age*****:** The difference between calendar years and the year seed articles were published is captured using subfield age dummies, included for each subfield up to the end of 2019. This approach controls for trends related to the age of the subfield without imposing a functional form.

B. *Subfield Growth Pattern*

- ***Cumulative Subfield Size*****:** The total number of articles published in a subfield up to a given year. This measure controls for the potential impact of a subfield's size on citation dynamics. We apply a logarithmic transformation to address skewness for robust statistical comparisons between subfields of varying sizes.

- ***Two Rolling-Years Marginal Growth*****:** The proportion of articles published in the current year and the previous year, divided by cumulative subfield size. This metric provides a normalized indicator of how actively a subfield is growing, shrinking, or remaining stagnant, adjusting for short-term fluctuations in publication activity that might affect outcomes of interest, such as citation dynamics.

C. Citation Dynamics

- ***Total Cumulative Citations:*** Aggregate citation counts that publications within a subfield have received up until a specified year. We include this variable to control for the overall academic impact of a subfield, which may influence the likelihood of sudden changes in citation patterns. A natural logarithmic transformation is applied to address skewness.

- ***Two-year Rolling Citation Counts***: Citations a subfield accumulates during the given year and the past year. We take the natural logarithm of the raw counts. This variable controls for the recent volume of citations, separate from long-term trends.

- ***Gini Coefficient of Citation Counts***: The Gini coefficient measures the degree of centralization in citation counts within a subfield. The coefficient ranges from 0 (where every article in a subfield receives the same number of citations) to 1 (where a single article receives all citations). Annually, we compute the Gini coefficients for 1) *Total Cumulative Citations* and 2) *Two-year Rolling Citation* to control for the potential impact of citation concentration.

*D. Other Controls*

- ***Article Retraction Notification***: Indicator variable that switches from 0 to 1 once a

retraction notification is observed in a subfield. It controls for the potential impact that experiencing a retraction event in the subfield level might have on overall attention the subfield receives.

- ***After Death (of Superstar Scientists)***: Indicator variable that switches from 0 to 1 with the death of superstar scientists (Supplementary Information 3.1). This attempts to capture any residual temporal effects of star death on citation dynamics.[26]

- ***After Death (of Superstar Scientists) * Subfields Associated with Premature Death of Superstar Scientists:*** The first term is as previously described. The latter is an indicator variable that differentiates subfields associated with the premature deaths of elite scientists from those that are not. This controls for the impact of the sudden death of star scientists on citation dynamics, reflecting how the data set was originally constructed and the finding that star death is *positively* associated with increases in subfield citation.[26]

- ***Calendar Year Fixed-Effect*****:** Year dummies to account for potential effects of calendar year from 1970 and 2019. We include this to ensure that any time-specific external influences are controlled across all subfields.

- ***Strata ID***: 3,076 “strata” IDs identified from the subfields associated with publications of prematurely deceased superstars(Azoulay et al. 2019). These IDs are assigned to comparable “within strata” subfields not experiencing a loss of star scientists. These comparable subfields are matched with those experiencing a star death based on key metrics such as 1) publication years, 2) team sizes, 3) ages of associated scientists, and 4) long-run citation impact, as detailed in Supplementary Information S3.1.

**Authors Contributions:**

Conceptualization: DK, RD, JR, JE

Methodology: DK, JE

Visualization: DK

Funding acquisition: RD, JE

Writing – original draft: DK, JE

Writing – review & editing: DK, RD, JR, JE

**Competing Interests:** The authors declare no competing interests.

**Data availability:** This work uses PubMed Knowledge Graph *(13)* and the replication data from Azoulay and colleagues *(14)*. They are available publicly.

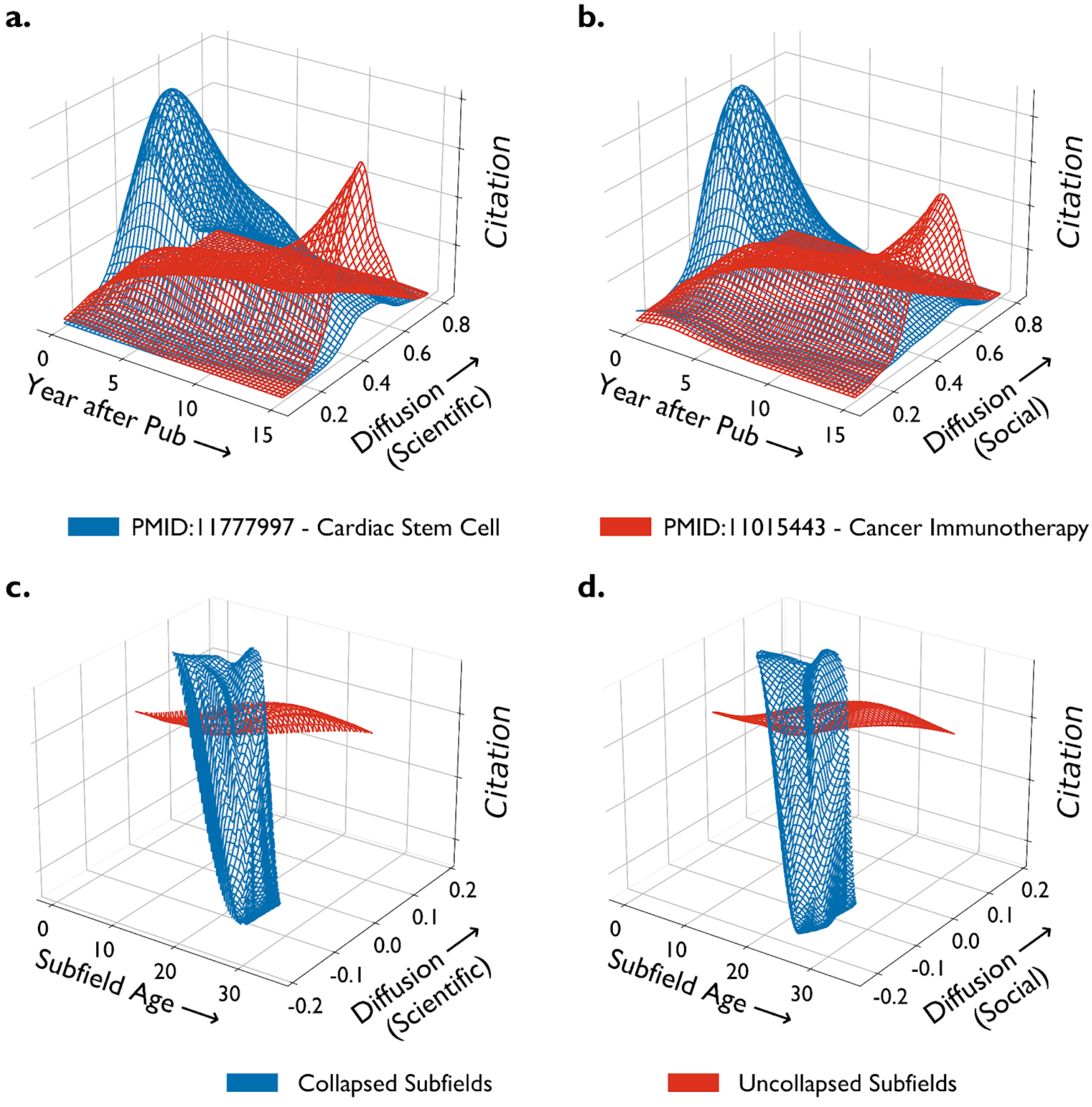

**Fig. 1: Representation of different diffusion levels and contrasting diffusion trajectories.** Panels **a** and **b** illustrate 3D kernel density plots of diffusion indices and citations for PMID 11777997 (Cardiac Stem Cell) and PMID 11015443 (Cancer Immunotherapy) in scientific and social spaces, respectively. Publication years associated with each article are aligned to zero for comparison. Annual diffusion indices and citation counts are computed using a two-year rolling average. Panels **c** and **d** show kernel density plots based on average diffusion indices and citations, standardized within subfield ages. These plots contrast subfields that experienced collapse below the 0.5% threshold (blue) with those that did not (red), across scientific and social spaces respectively.

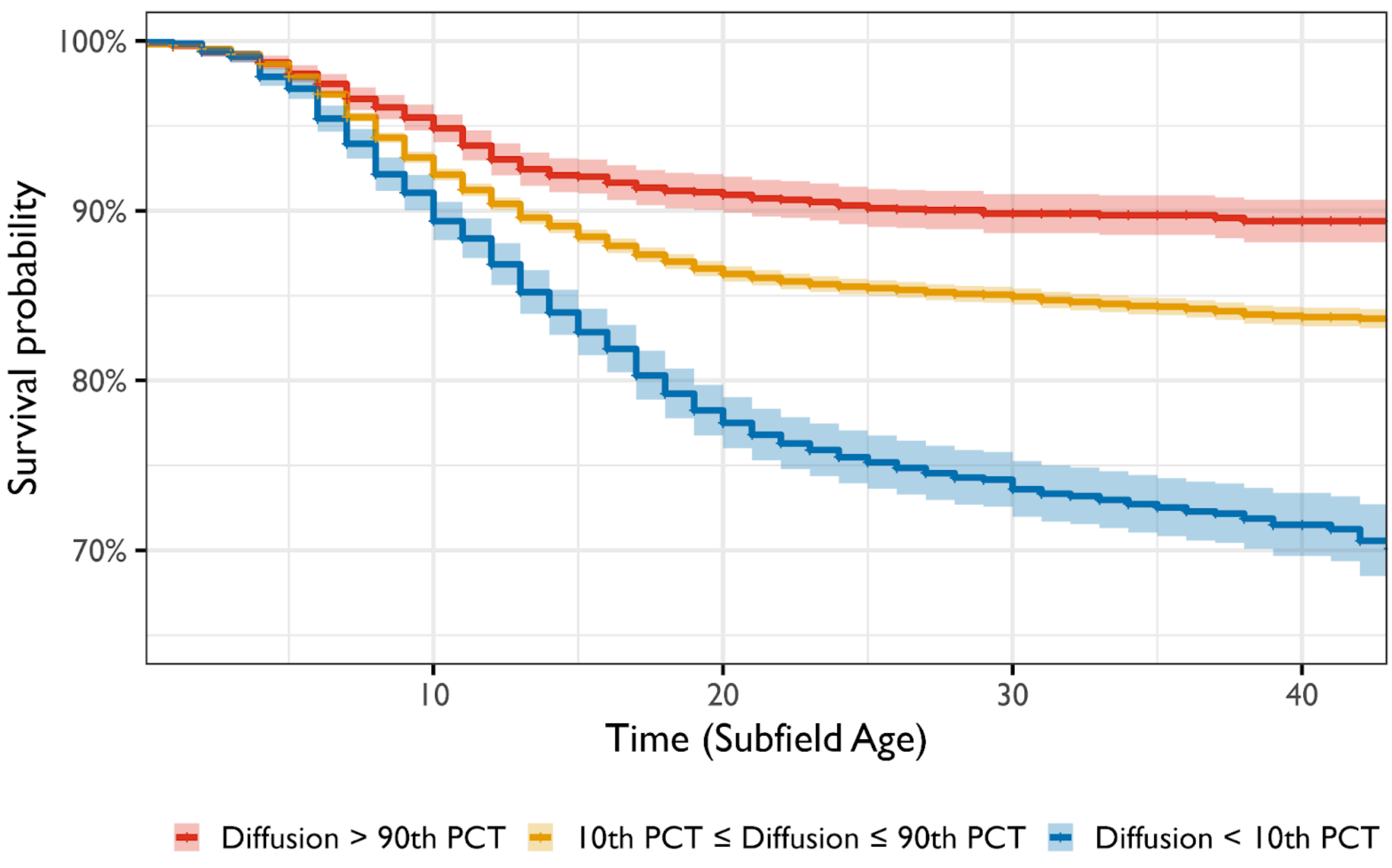

**Fig. 2: Survival probability against bubble bursting as a function of knowledge diffusion in social space.** Events are defined as a sharp decline of 2 year citation counts at the subfield level with 0.5% cutoff (Method). Survival refers to the converse, i.e., not experiencing a subfield-level extreme deflationary event. Subfield ages are set to 0 in the year when the focal seed article spanning a subfield was published. Diffusion percentile is ranked within calendar years and subfield ages. Bands depict 95% confidence intervals.

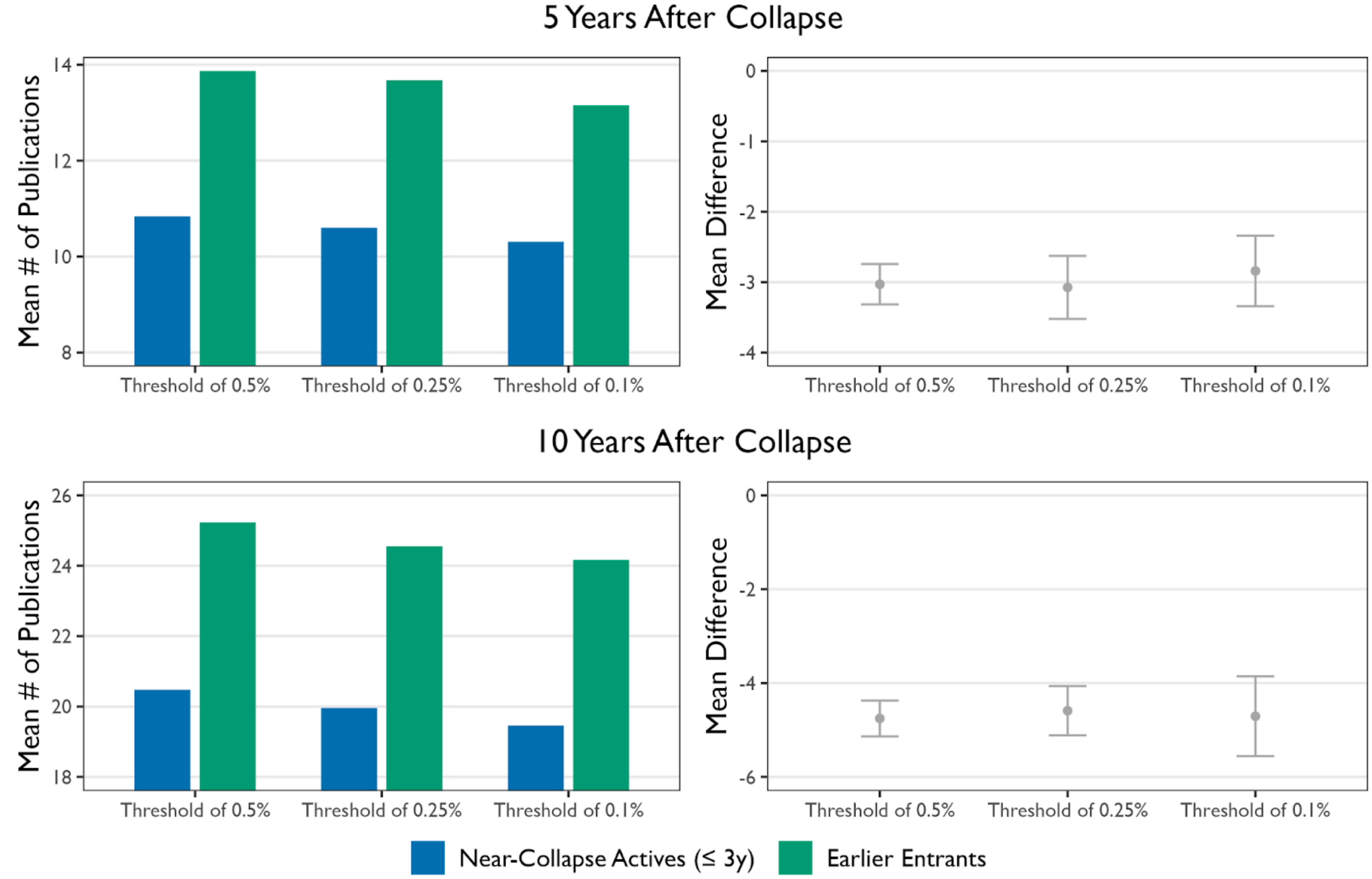

**Fig. 3: Comparing author productivity in collapsed subfields 5 and 10 years post-collapse.** The error bars represent the 95% confidence intervals for the mean differences in average publication numbers. Comparisons are drawn between authors who entered the field early and those active near the collapse, based on paired t-tests.

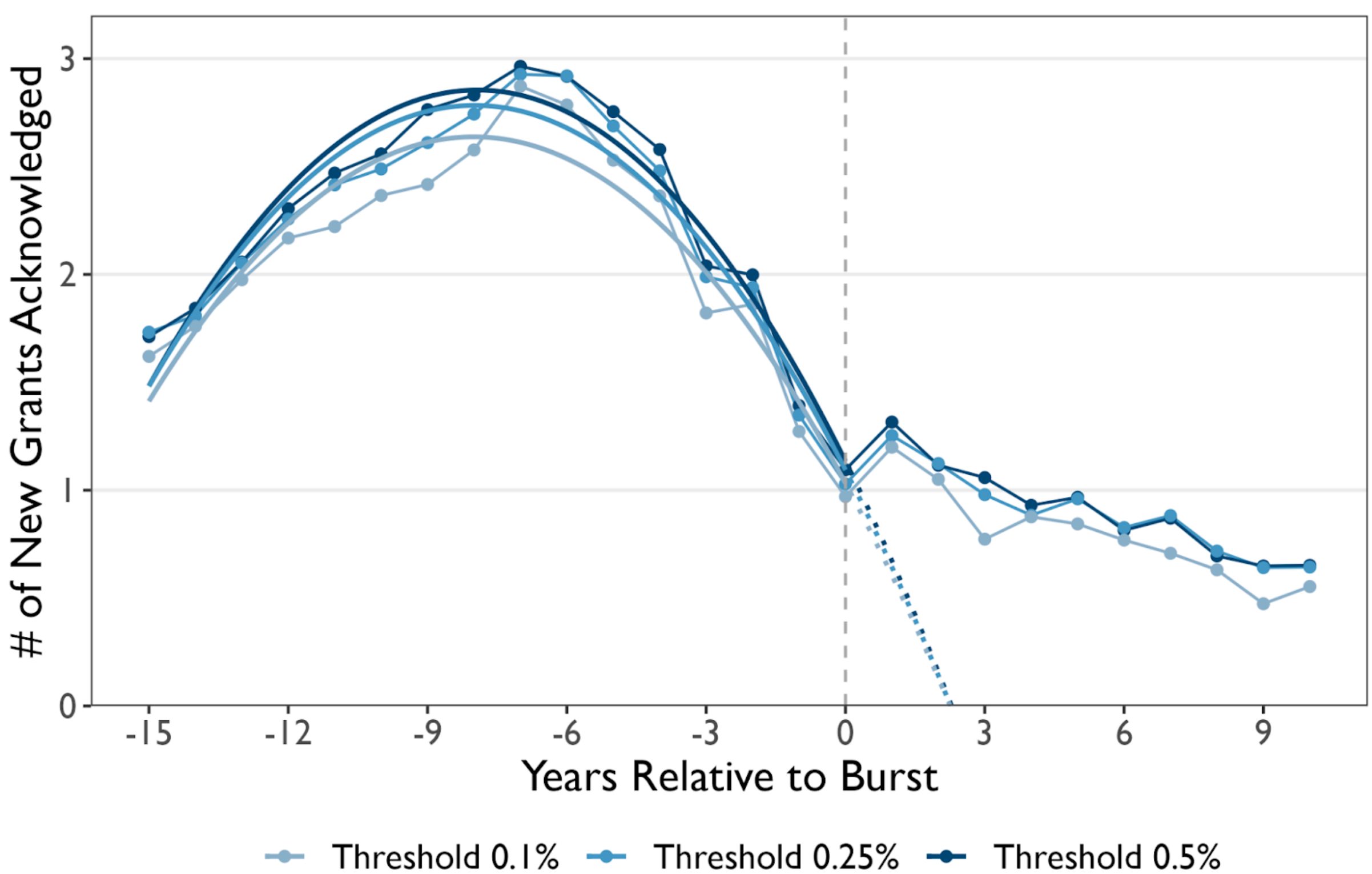

**Fig. 4: The average number of new grants acknowledged in collapsed subfields by years relative to burst.** The quadratic fit is applied to data from years -15 to 0 relative to the burst year, with dotted lines representing extrapolations starting from year 0 onwards.

| Dependent Variable | *Substantial Decline of Citations* | | | | |
|---|---|---|---|---|---|
| | **Estimate** | **Std. Error** | ***t*** | ***p*-value** | **95% C.I.** |
| ***Diffusion*** | | | | | |
| Scientific Space | -0.204 | 0.039 | -5.228 | < 0.001 | [-0.280, -0.128] |
| Social Space | -0.277 | 0.049 | -5.598 | < 0.001 | [-0.373, -0.181] |
| ***Subfield Growth Pattern*** | | | | | |
| Cum. Subfield Size (logged) | 0.499 | 0.084 | 5.923 | < 0.001 | [0.334, 0.664] |
| 2-year Subfield Marginal Growth | -0.217 | 0.010 | -21.543 | < 0.001 | [-0.237, -0.197] |
| ***Citation Dynamics*** | | | | | |
| Cum. Citations (logged) | -2.472 | 0.164 | -15.003 | < 0.001 | [-2.793, -2.151] |
| 2-year Citations (logged) | 2.772 | 0.146 | 18.942 | < 0.001 | [2.486, 3.058] |
| Gini Coef. of Cum. Citations | 0.012 | 0.006 | 1.843 | 0.065 | [0.000, 0.024] |
| Gini Coef. of 2-year Citations | -0.026 | 0.006 | -4.710 | 0.001 | [-0.038, -0.014] |
| ***Other Controls*** | | | | | |
| Retraction Notice Published | 0.062 | 0.199 | 0.313 | 0.754 | [-0.328, 0.452] |
| After Death | 0.152 | 0.129 | 1.186 | 0.236 | [-0.101, 0.405] |
| After Death * Superstar Death | -0.138 | 0.086 | -1.601 | 0.109 | [-0.307, 0.031] |
| Log-Likelihood | -26,289.5 | | | | |
| **Total Observations** | 1,313,433 | | | | |

**Table 1**: Model estimates with the bottom 0.5% cutoff for citation differences in the two-year rolling period. Coefficients for fixed effects of field age, calendar year, and strata ID dummies are omitted. Variables under *Knowledge Diffusion*, *Subfield Growth Pattern*, *and Citation Dynamics* are all one-year lagged. The diffusion indices are standardized within field ages and calendar years across 28,504 subfields. Standard errors are clustered with strata ID and calendar years.

**Acknowledgement**: We express gratitude for funding from the Fetzer Franklin Fund in association with the 2019 MetaScience Symposium (RD, JE, DK), the Air Force Office of Scientific Research (AFOSR: FA9550-19-1-0354 and FA9550-15-1-0162) (JE, DK), and the National Science Foundation (NSF: 1829366 and 1800956) (JE, DK). The funders have/had no role in study design, data collection and analysis, decision to publish or preparation of the manuscript. This work was completed in part with resources provided by the University of Chicago's Research Computing Center. We also appreciate the support from Jian Xu and Ying Ding in facilitating access to PubMed Knowledge Graph.

## Extended Data

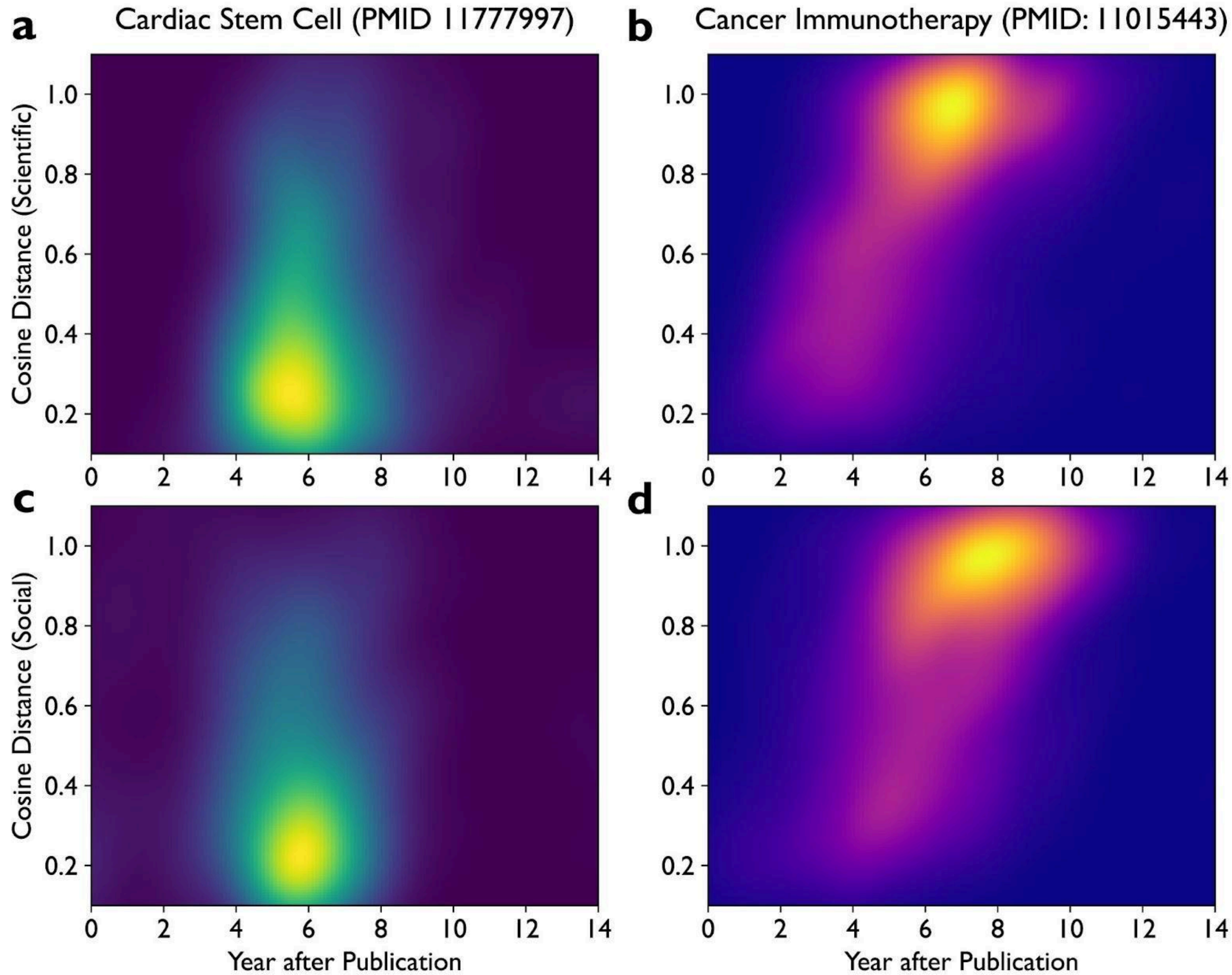

**Extended Data Fig. 1: Complementary 2D heatmaps for the upper panels of Fig. 1.** Values are derived from the identical kernel density estimations graphed in Fig. 1 for the distribution of diffusion indices in scientific (panels **a** and **b**) and social spaces (panels **c** and **d**), respectively.

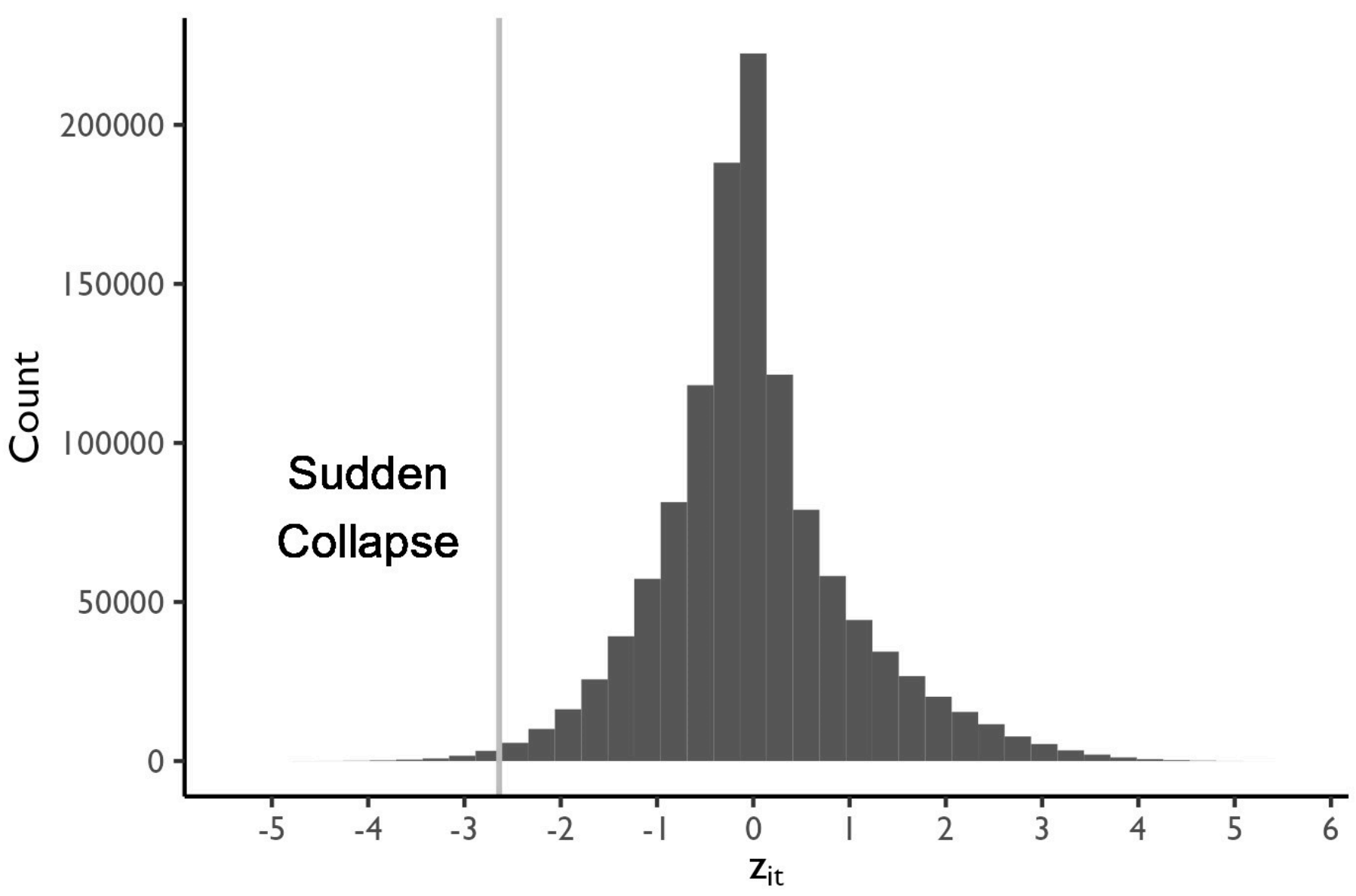

**Extended Data Fig. 2: Distribution of $z_{i,t}$ from 28,504 subfields**. The cutoff value for bubble burst here is set to -2.64, the bottom 0.5% percentile. The range of $z_{i,t}$ is [-5.2, 5.52].

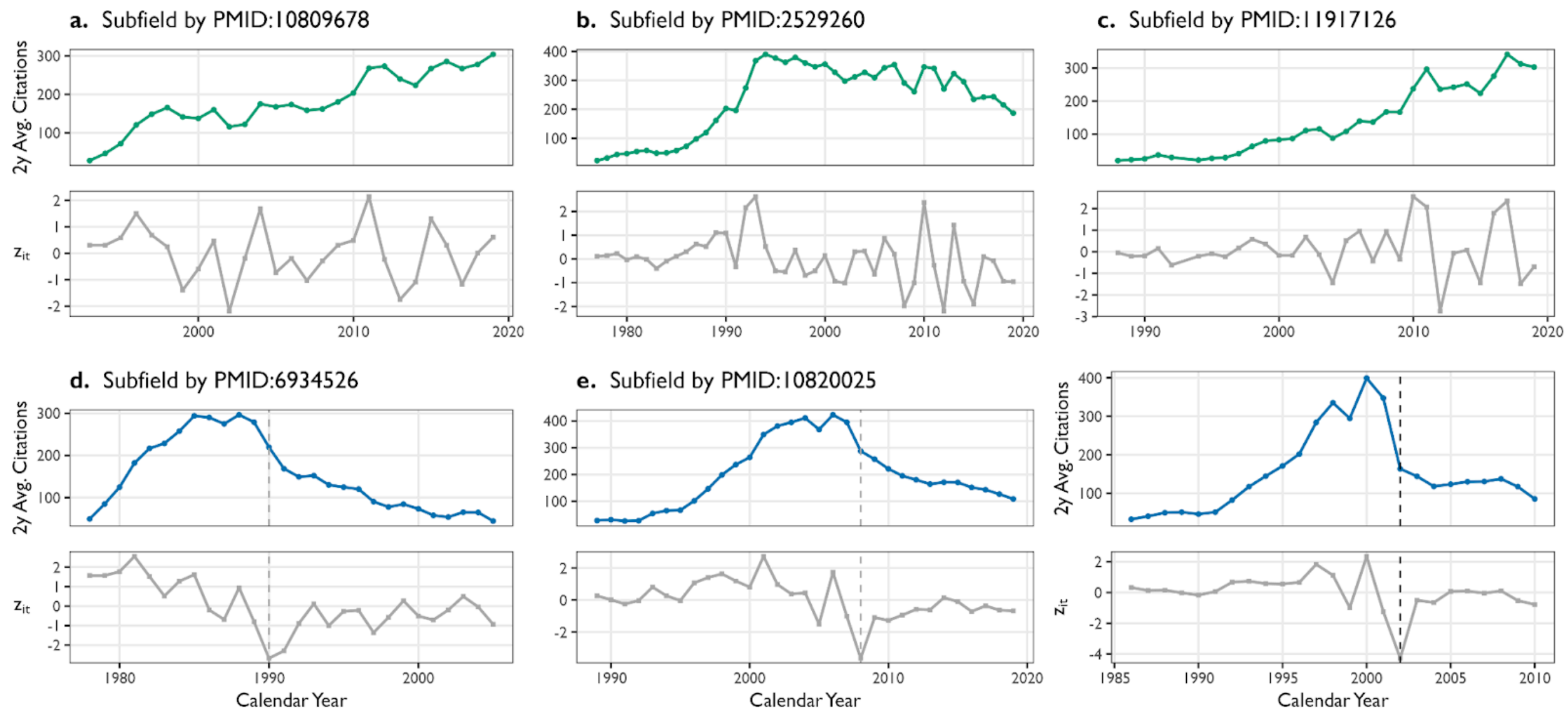

**Extended Data Fig. 3: Six examples of subfields.** Annual citation counts aggregated at the subfield level, using forward citations to related publications. Top panels (a, b, c): Subfields represented by three PMIDs, illustrating cases without bubble bursting events. Bottom panels (d, e, f): Subfields that experienced bubble bursting, corresponding to the cutoffs closest to the 0.5%, 0.25%, and 0.1% thresholds of $z_{i,t}$ value.

| | Threshold | Estimate | t | *p*-value (*d.f.*) | 95% C.I. |
|---|---|---|---|---|---|
| **5 Years** | 0.5% | -3.030 | -20.67 | < 0.001 (3,910) | [-3.317, -2.743] |
| | 0.25% | -3.075 | -13.48 | < 0.001 (1,983) | [-3.522, -2.628] |
| | 0.1% | -2.841 | -11.13 | < 0.001 (773) | [-3.342, -2.340] |
| **10 Years** | 0.5% | -4.754 | -24.45 | < 0.001 (3,605) | [-5.136, -4.373] |
| | 0.25% | -4.590 | -17.17 | < 0.001 (1,818) | [-5.114, -4.066] |
| | 0.1% | -4.707 | -10.87 | < 0.001 (711) | [-5.558, -3.857] |

**Extended Data Table 1**: **Pairwise *t*-test comparing average productivity differences between near-collapse active scientists (≤2 years before collapse) and early entrants.** Subfields that collapsed after 2015 were excluded from the 5-year productivity comparison. Likewise, for the 10-year productivity, only subfields that collapsed on or before 2011 were included, considering the observation windows.

| Threshold | % of Subfields with New Grants Acknowledged After Collapse | Mean | Q1 | Median | Q3 |
|---|---|---|---|---|---|
| 0.5% | 83.12% | 11.8 | 2 | 6 | 16 |
| 0.25% | 82.93% | 11.3 | 2 | 6 | 15 |
| 0.1% | 81.70% | 10.1 | 1 | 6 | 13 |

**Extended Data Table 2: Proportion of subfields with newly acknowledged grants after collapse, and the Mean, 1st Quantile, Median, and 3rd Quartile of the number of new grants post-collapse.** Subfields that collapsed after 2015 were excluded from the 5-year productivity comparison. For the 10-year productivity analysis, only subfields that collapsed on or before 2011 were included, in consideration of the observation windows.

# SUPPLEMENTARY INFORMATION FOR

# Limited Diffusion of Scientific Knowledge Forecasts Collapse

## S1. Supplementary Figures and Tables

**Table S1.1:** Model Estimates with the Bottom 0.5% Cutoff, with and without Controls.

| **Dependent Variable** | | | *Substantial Decline of Citation* | | |
|---|---|---|---|---|---|
| | **Model 1** | **Model 2** | **Model 3** | **Model 4** | **Model 5** |
| ***Knowledge Diffusion*** | | | | | |
| Scientific Space | -0.217***<br>[-0.293, -0.141]<br>($p < 0.001$) | -0.271***<br>[-0.350, -0.192]<br>($p < 0.001$) | -0.137***<br>[-0.207, -0.068]<br>($p < 0.001$) | -0.138***<br>[-0.208, -0.069]<br>($p < 0.001$) | -0.204***<br>[-0.28, -0.127]<br>($p < 0.001$) |
| Social Space | -0.218***<br>[-0.303, -0.134]<br>($p < 0.001$) | -0.277***<br>[-0.369, -0.184]<br>($p < 0.001$) | -0.248***<br>[-0.339, -0.158]<br>($p < 0.001$) | -0.245***<br>[-0.336, -0.155]<br>( $p < 0.001$) | -0.277***<br>[-0.374, -0.180]<br>($p < 0.001$) |
| ***Subfield Growth*** | | | | | |
| Cum. Subfield Size (logged) | | 0.240***<br>[0.168, 0.313]<br>($p < 0.001$) | 0.726***<br>[0.602, 0.851]<br>($p < 0.001$) | 0.739***<br>[0.613, 0.864]<br>($p < 0.001$) | 0.499***<br>[0.334, 0.664]<br>($p < 0.001$) |
| 2-year Subfield Growth | | -0.116***<br>[-0.130, -0.103]<br>($p < 0.001$) | -0.197***<br>[-0.214, -0.179]<br>($p < 0.001$) | -0.197***<br>[-0.215, -0.180]<br>($p < 0.001$) | -0.217***<br>[-0.237, -0.198]<br>($p < 0.001$) |
| ***Citation Dynamics*** | | | | | |
| Cum. Citations (logged) | | | -2.437***<br>[-2.675, -2.199]<br>($p < 0.001$) | -2.458***<br>[-2.700, -2.215]<br>($p < 0.001$) | -2.472***<br>[-2.794, -2.149]<br>($p < 0.001$) |
| 2-year Citations (logged) | | | 2.378***<br>[2.128, 2.629]<br>($p < 0.001$ | 2.386***<br>[2.134, 2.638]<br>($p < 0.001$) | 2.772***<br>[2.485, 3.058]<br>($p < 0.001$) |
| Gini Coef. of Cum. Citation | | | 0.011<br>[0.000, 0.022]<br>($p = 0.050$) | 0.012*<br>[0.001, 0.023]<br>($p = 0.04$) | 0.012<br>[-0.001, 0.024]<br>($p = 0.065$) |
| Gini Coef. of 2-year Citation | | | -0.018***<br>[-0.028, -0.009]<br>($p < 0.001$) | -0.018***<br>[-0.028, -0.009]<br>($p < 0.001$) | -0.026***<br>[-0.037, -0.015]<br>($p < 0.001$) |
| ***Other Controls*** | | | | | |
| Retraction Notice Published | | | | -0.035<br>[-0.405, 0.335]<br>($p = 0.853$) | 0.062<br>[-0.328, 0.453]<br>($p = 0.754$) |
| After Death | | | | 0.288<br>[0.012, 0.563]<br>($p = 0.041$) | 0.152<br>[-0.100, 0.404]<br>($p = 0.236$) |
| After Death * Superstar Death | | | | -0.186**<br>[-0.322, -0.05]<br>($p = 0.007$) | -0.138<br>[-0.306, 0.031]<br>($p = 0.109$) |
| ***Fixed Effects*** | | | | | |
| Calendar Year | N | N | N | N | Y |
| Strata ID | N | N | N | N | Y |
| **Log-Likelihood** | -31,686.1 | -30,832.3 | -28,989.6 | -28,978.4 | -26,289.5 |
| **Total Observations** | | | 1,313,433 | | |

**Note:** *Knowledge Diffusion*, *Subfield Growth Pattern*, *and Citation Dynamics* are all one-year lagged. The diffusion indices are standardized within field ages and calendar years across 28,504 subfields. The 95% confidence intervals inside brackets are computed based on the standard errors clustered at strata ID and calendar years. Coefficients for field ages are omitted. * $p < .05$; ** $p < .01$; *** $p < .001$ (two-tailed)

**Table S1.2:** Model Estimates with the Bottom 0.25% Cutoff, with and without Controls.

| **Dependent Variable** | *Substantial Decline of Citation* | | | | |
|---|---|---|---|---|---|
| | **Model 1** | **Model 2** | **Model 3** | **Model 4** | **Model 5** |
| ***Knowledge Diffusion*** | | | | | |
| Scientific Space | -0.234***<br>[-0.323, -0.145]<br>($p < 0.001$) | -0.289***<br>[-0.382, -0.196]<br>($p < 0.001$) | -0.140***<br>[-0.220, -0.061]<br>($p < 0.001$) | -0.141***<br>[-0.221, -0.061]<br>($p < 0.001$) | -0.221***<br>[-0.302, -0.139]<br>($p < 0.001$) |
| Social Space | -0.260***<br>[-0.363, -0.156]<br>($p < 0.001$) | -0.323***<br>[-0.438, -0.209]<br>($p < 0.001$) | -0.289***<br>[-0.400, -0.178]<br>($p < 0.001$) | -0.285***<br>[-0.396, -0.174]<br>($p < 0.001$) | -0.313***<br>[-0.431, -0.196]<br>($p < 0.001$) |
| ***Subfield Growth*** | | | | | |
| Cum. Subfield Size (logged) | | 0.276***<br>[0.190, 0.361]<br>($p < 0.001$) | 0.830***<br>[0.681, 0.979]<br>($p < 0.001$) | 0.846***<br>[0.697, 0.995]<br>($p < 0.001$) | 0.649***<br>[0.408, 0.890]<br>($p < 0.001$) |
| 2-year Subfield Growth | | -0.126***<br>[-0.145, -0.107]<br>($p < 0.001$) | -0.211***<br>[-0.234, -0.189]<br>($p < 0.001$) | -0.212***<br>[-0.234, -0.19]<br>($p < 0.001$) | -0.240***<br>[-0.266, -0.214]<br>($p < 0.001$) |
| ***Citation Dynamics*** | | | | | |
| Cum. Citations (logged) | | | -2.624***<br>[-2.900, -2.347]<br>($p < 0.001$) | -2.648***<br>[-2.927, -2.370]<br>($p < 0.001$) | -2.739***<br>[-3.098, -2.380]<br>($p < 0.001$) |
| 2-year Citations (logged) | | | 2.545***<br>[2.253, 2.836]<br>($p < 0.001$) | 2.555***<br>[2.263, 2.847]<br>($p < 0.001$) | 2.976***<br>[2.652, 3.301]<br>($p < 0.001$) |
| Gini Coef. of Cum. Citation | | | 0.011<br>[-0.005, 0.028]<br>($p = 0.172$) | 0.012<br>[-0.004, 0.029]<br>($p = 0.153$) | 0.011<br>[-0.007, 0.029]<br>($p = 0.241$) |
| Gini Coef. of 2-year Citation | | | -0.021**<br>[-0.035, -0.008]<br>($p = 0.002$) | -0.021**<br>[-0.035, -0.008]<br>($p = 0.002$) | -0.029***<br>[-0.044, -0.013]<br>($p < 0.001$) |
| ***Other Controls*** | | | | | |
| Retraction Notice Published | | | | -0.166<br>[-0.653, 0.321]<br>($p = 0.503$) | -0.069<br>[-0.594, 0.456]<br>($p = 0.796$) |
| After Death | | | | 0.404*<br>[0.026, 0.782]<br>($p = 0.036$) | 0.263<br>[-0.078, 0.604]<br>($p = 0.130$) |
| After Death * Superstar Death | | | | -0.205*<br>[-0.380, -0.031]<br>($p = 0.021$) | -0.150<br>[-0.371, 0.071]<br>($p = 0.182$) |
| ***Fixed Effects*** | | | | | |
| Calendar Year | N | N | N | N | Y |
| Strata ID | N | N | N | N | Y |
| **Log-Likelihood** | -18,129.3 | -17,635.5 | -16,532.5 | -16,523.1 | -14,527.7 |
| **Total Observations** | | | 1,366,970 | | |

**Note:** *Knowledge Diffusion*, *Subfield Growth Pattern*, *and Citation Dynamics* are all one-year lagged. The diffusion indices are standardized within field ages and calendar years across 28,504 subfields. The 95% confidence intervals inside brackets are computed based on the standard errors clustered at strata ID and calendar years. Coefficients for field ages are omitted. * $p < .05$; ** $p < .01$; *** $p < .001$ (two-tailed)

**Table S1.3:** Model Estimates with the Bottom 0.1% Cutoff, with and without Controls.

| **Dependent Variable** | *Substantial Decline of Citation* | | | | |
|---|---|---|---|---|---|
| | **Model 1** | **Model 2** | **Model 3** | **Model 4** | **Model 5** |
| ***Knowledge Diffusion*** | | | | | |
| Scientific Space | -0.209**<br>[-0.342, -0.076]<br>($p = 0.002$) | -0.263***<br>[-0.404, -0.122]<br>($p < 0.001$) | -0.105<br>[-0.228, 0.017]<br>($p = 0.091$) | -0.107<br>[-0.23, 0.016]<br>($p = 0.090$) | -0.179**<br>[-0.297, -0.060]<br>($p = 0.003$) |
| Social Space | -0.344***<br>[-0.506, -0.183]<br>($p < 0.001$) | -0.419***<br>[-0.603, -0.235]<br>($p < 0.001$) | -0.376***<br>[-0.550, -0.203]<br>($p < 0.001$) | -0.371***<br>[-0.544, -0.197]<br>($p < 0.001$) | -0.435***<br>[-0.642, -0.229]<br>($p < 0.001$) |
| ***Subfield Growth*** | | | | | |
| Cum. Subfield Size (logged) | | 0.323***<br>[0.197, 0.448]<br>($p < 0.001$) | 0.967***<br>[0.791, 1.142]<br>($p < 0.001$) | 0.992***<br>[0.814, 1.169]<br>($p < 0.001$) | 0.980***<br>[0.683, 1.278]<br>($p < 0.001$) |
| 2-year Subfield Growth | | -0.139***<br>[-0.171, -0.107]<br>($p < 0.001$) | -0.227***<br>[-0.258, -0.196]<br>($p < 0.001$) | -0.228***<br>[-0.26, -0.197]<br>($p < 0.001$) | -0.271***<br>[-0.308, -0.234]<br>($p < 0.001$) |
| ***Citation Dynamics*** | | | | | |
| Cum. Citations (logged) | | | -2.751***<br>[-3.100, -2.401]<br>($p < 0.001$) | -2.793***<br>[-3.147, -2.44]<br>($p < 0.001$) | -3.117***<br>[-3.545, -2.689]<br>($p < 0.001$) |
| 2-year Citations (logged) | | | 2.612***<br>[2.232, 2.992]<br>($p < 0.001$) | 2.626***<br>[2.246, 3.006]<br>($p < 0.001$) | 3.111***<br>[2.684, 3.538]<br>($p < 0.001$) |
| Gini Coef. of Cum. Citation | | | 0.011<br>[-0.022, 0.043]<br>($p = 0.515$) | 0.012<br>[-0.021, 0.044]<br>($p = 0.482$) | 0.017<br>[-0.015, 0.05]<br>($p = 0.298$) |
| Gini Coef. of 2-year Citation | | | -0.019<br>[-0.047, 0.009]<br>($p = 0.178$) | -0.019<br>[-0.047, 0.009]<br>($p = 0.179$) | -0.025<br>[-0.058, 0.008]<br>($p = 0.131$) |
| ***Other Controls*** | | | | | |
| Retraction Notice Published | | | | 0.224<br>[-0.469, 0.916]<br>($p = 0.527$) | 0.385<br>[-0.377, 1.147]<br>($p = 0.322$) |
| After Death | | | | 0.764***<br>[0.318, 1.211]<br>($p < 0.001$) | 0.556*<br>[0.073, 1.039]<br>($p = 0.024$) |
| After Death * Superstar Death | | | | -0.289<br>[-0.584, 0.005]<br>($p = 0.054$) | -0.163<br>[-0.574, 0.249]<br>($p = 0.438$) |
| ***Fixed Effects*** | | | | | |
| Calendar Year | N | N | N | N | Y |
| Strata ID | N | N | N | N | Y |
| **Log-Likelihood** | -8,107.5 | -7,879.7 | -7,405.3 | -7,395.6 | -6,010.4 |
| **Total Observations** | | | 1,401,037 | | |

**Note:** *Knowledge Diffusion*, *Subfield Growth Pattern*, *and Citation Dynamics* are all one-year lagged. The diffusion indices are standardized within field ages and calendar years across 28,504 subfields. The 95% confidence intervals inside brackets are computed based on the standard errors clustered at strata ID and calendar years. Coefficients for field ages are omitted. * $p < .05$; ** $p < .01$; *** $p < .001$ (two-tailed)

## S2. Measuring Knowledge Diffusion Through Document Embedding Spaces

We train vector representation models for biomedical science publications from PKG 2020 to locate positions of scientific publications based on their contents and to measure the similarity/distance between papers linked through citations. We adapt the Doc2vec model[1], a variant of the Word2vec model[2,3], which was initially developed to produce dense vector representations for documents or paragraphs from the words that compose them. Word embedding models generate a high-dimensional vector space in which geometrically proximate word vectors correspond to words that frequently share local linguistic contexts in the training data[2–4]. This approach has previously been extended to generate representational vectors for entities connected in networks by substituting connections among entities as shared contexts[5,6].

We consider that a research article can be characterized by 1) a list of MeSH terms and 2) researchers authoring it. Accordingly, we build two separate representational vector spaces — "scientific space" and "social space." We employ the Python Gensim package (version 4.0)[7] to train our vector representations. We specifically use the Distributed Bag of Words (DBOW) model, analogous to the skip-gram model from the Word2vec framework, to train document vectors and constituting elements (MeSH terms and author IDs) simultaneously. This approach enables us to conduct document retrieval tasks using the vector representations of MeSH terms and author IDs to validate the resulting spaces. Detailed implementation procedures are as follows.

### *S2.1 Scientific Space from MeSH Descriptors*

We posit the MeSH terms as constituting words to build a "scientific space" for the biomedical literature. Because nominal terminologies are subject to change, we use MeSH terms' unique IDs from the National Library of Medicine. For instance, a MeSH descriptor, *Mesenchymal Stem Cells* (Descriptor ID: *D059630*), was indexed as *Mesenchymal Stromal Cells* from 2012 to 2018. However, it began to be reindexed as *Mesenchymal stem cells* in 2019, while its uniquely assigned descriptor ID, *D059630*, remains the same.

**MeSH terms**

- Antibodies, Monoclonal / therapeutic use
- Antineoplastic Agents / metabolism
- Antineoplastic Agents / pharmacology*
- Humans
- Immunologic Factors / therapeutic use*
- Immunotherapy*
- Neoplasms / therapy*
- Programmed Cell Death 1 Receptor / therapeutic use*
- Signal Transduction

**Fig. S2.1: MeSH terms assigned to PMID 28376884**

When a MeSH qualifier is attached to a MeSH descriptor, we consider both a descriptor with a qualifier and without it. Note that Fig. S2.1 displays MeSH terms assigned to "Cancer immunotherapies targeting the PD-1 signaling pathway" (PMID 28376884), published in the *Journal of Biomedical Science* in 2017, authored by Iwai, Hamanishi, Chamoto, and Honjo[8]. The second term, *Antineoplastic Agents / metabolism*, can be broken down into the primary MeSH descriptor, *Antineoplastic Agents*, and the qualifier, *metabolism*, narrowing down the scope. The third term, _Antineoplastic Agents / pharmacology_*, also has a qualifier, *pharmacology*. (The asterisk denotes that the given term is a major topic of the publication.) For this case, we include 1) *Antineoplastic Agents,* 2) *Antineoplastic Agents / metabolism,* and 3) *Antineoplastic Agents / pharmacology* for our model training. We do this to reflect that PubMed search queries using only MeSH terms (without qualifiers), *Antineoplastic Agents*, for this case, capture publications like PMID 28376884. We exclude the asterisks for the same reason, taking into consideration co-searchability. As a result, the final list of MeSH terms fed into the training process for PMID 28376884 is A*ntibodies, Monoclonal*, *Antibodies*, *Monoclonal / therapeutic use*, *Antineoplastic Agents*, *Antineoplastic Agents / metabolism*, *Antineoplastic Agents / pharmacology*, *Humans*, *Immunologic Factors*, *Immunologic Factors / therapeutic use*, *Immunotherapy*, *Neoplasms*, *Neoplasms / therapy*, *Programmed Cell Death 1 Receptor*, *Programmed Cell Death 1 Receptor / therapeutic use*, *Signal Transduction*.

With these MeSH combinations, we train 100-dimensional vectors for 26,666,615 PMIDs and 303,492 MeSH combinations that appear at least ten times with 100 training epochs. The mean number of MeSH terms (after the procedure detailed above) per PMID from our dataset is 16.34 (std=9.04). However, we set the sliding window size that defines the boundary of the training context as 110, the maximum number from the data, to ensure that each training instance includes all the other MeSH combinations on a given article without splitting them up by imposing arbitrary contexts.

We validate the resulting vector representations by attempting to retrieve resulting publication vectors using MeSH combination vectors across 20 random samples, each containing 1,000 publications. We first take the vectors of MeSH terms assigned to each publication, infer the position of a document combining the MeSH terms, and check its proximity to the original vector representation of the article containing those MeSH terms. It is, for instance, a test to see if we can retrieve PMID 28376884 in Fig. S2.1 by inferring the position of a document combining the vectors of MeSH terms assigned to it. Because it is impossible to differentiate publications with the same set of MeSH terms with this model, we consider the 1, 5, and 10 most similar documents from the inferred vector, using cosine similarity. We find that it is possible to retrieve the target PMIDs with the rate of 92.48% (sd= .81), 96.14% (sd=.59), and 97.18% (sd=.52) from the top 1, 5, and 10 most similar documents, respectively, which suggests documents sharing MeSH terms are located close together in the 100-dimensional embedding space.

An advantage of using this Doc2vec model is that it reflects the high-order proximity of constituting words beyond their direct co-occurrence in a context. Consider two documents, PMID 23142641, a review article titled “Challenges measuring cardiomyocyte renewal,” published in 2013[9], and PMID 11287958, an original research article, “Bone marrow cells regenerate infarcted myocardium,” published in 2001[10]. The former review article cited the latter article. A simple but popular similarity metric would be the Jaccard coefficient ranging from 0 to 1, computed by dividing the number of MeSH terms that two articles share by the size of the union set of all MeSH terms assigned to the two publications.

The MeSH terms assigned to PMID 23142641 are _Animals; Bromodeoxyuridine; Cell Differentiation; Cell Nucleus / metaboli`m; Cell Nucleus / ultrastructure; Cell Proliferation; Cell Tracking; Genes, Reporter; Integrases; Mice; Mice, Transgenic; Myocardium / cytology*;_

*Myocardium / metabolism; Myocytes, Cardiac / cytology*; Myocytes, Cardiac / metabolism; Regeneration; Stem Cells / cytology*; Stem Cells / metabolism; beta-Galactosidase*.

The MeSH terms assigned to PMID 11287958 are as follows: *Animals; Bone Marrow Transplantation*; Cell Differentiation; Connexin 43 / metabolism; DNA-Binding Proteins / metabolism; Female; Green Fluorescent Proteins; Ki-67 Antigen / metabolism; Luminescent Proteins / metabolism; MEF2 Transcription Factors; Male; Mice; Mice, Inbred C57BL; Mice, Transgenic; Myocardial Infarction / therapy*; Myocardium / cytology; Myocardium / pathology*; Myogenic Regulatory Factors; Proto-Oncogene Proteins c-kit / metabolism; Transcription Factors / metabolism*.

The Jaccard coefficient of the two publications based on the MeSH terms is .133 despite the close relationship between the two articles. However, the cosine similarity between the two documents on our trained model is .844, which better reflects the overall topic similarity between the two publications.

### *S2.2 Social Space with Disambiguated Author IDs*

Analogous to the content embedding space from MeSH terms, we also build a 100-dimensional social embedding space using Doc2vec, anchored by 8,359,189 disambiguated biomedical authors, within which we locate the vector space position of 28,329,992 PMIDs published by the end of 2019. In other words, we consider the author IDs as constituting document units. To inscribe the co-author information per publication, we included only authors that appeared more than once. The mean number of authors per publication from 28,329,992 PMIDs is 3.97 (std=5.01) with a median of 3. However, we set the window size for the training context as 2000 – arbitrarily larger than the maximum number of authors in the dataset – to include all author IDs in the training process for a given publication. We do this to ensure that the resulting article embedding model assigns similar vectors to articles co-authored by the same groups of overlapping co-authors who are directly or indirectly close in the social space of biomedical research collaboration. We trained our social embedding space using 100 epochs (or training iterations).

We validate the quality of vector representations in the same manner we did for the MeSH content space across 20 random samples of 1,000 publications each. We take the author vectors for each publication, infer the position of a hypothetical publication those authors could have written within the 100-dimensional embedding space, and check its proximity to the vector representation written by the same author(s). Considering the impossibility of distinguishing publications written by the same author(s), we also assess the 1, 5, 10, and 20 most similar PMIDs from the inferred vector using cosine similarity. The target PMIDs could be retrieved with the rate of 65.26% (sd= 1.73), 86.16% (sd=1.06), 90.27% (sd=0.74), 92.9% (sd=0.77) from the top 1, 5, 10, 20 most similar documents, respectively. The sharp increase in self-retrieval for relaxed conditions demonstrates that papers written by the same author(s) are contiguous in the resulting 100-dimensional social embedding space.

### *S2.3 Aggregated Temporal Pattern of Diffusion from Highly Cited Articles Published in 1980, 1990, 2000, 2010*

Here, we provide an aggregate-level description of how our diffusion indices temporally evolve using highly cited papers (top 5% percent in citation counts by the end of 2019) from four cohorts of research articles published in 1980, 1990, 2000, and 2010. We first make subsets of publications that the raw citation obtained by the end of 2019 fall over the 5% percentile in each cohort year (10,967 of 219,358 in 1970; 14,031 of 280,622 in 1980; 20,527 of 410,555 in 1990; 26,513 of 530,271 in 2000; 41,156 /823,129 in 2010), also accordingly extract cosine distances between the focal papers and citing papers measured in social and scientific space. With data from two rolling years, medians of cosine distances from two spaces each calendar year are computed. For example, the median cosine distance assigned to 1991 for the 1990 cohort is computed using all the citations observed in 1990 and 1991.

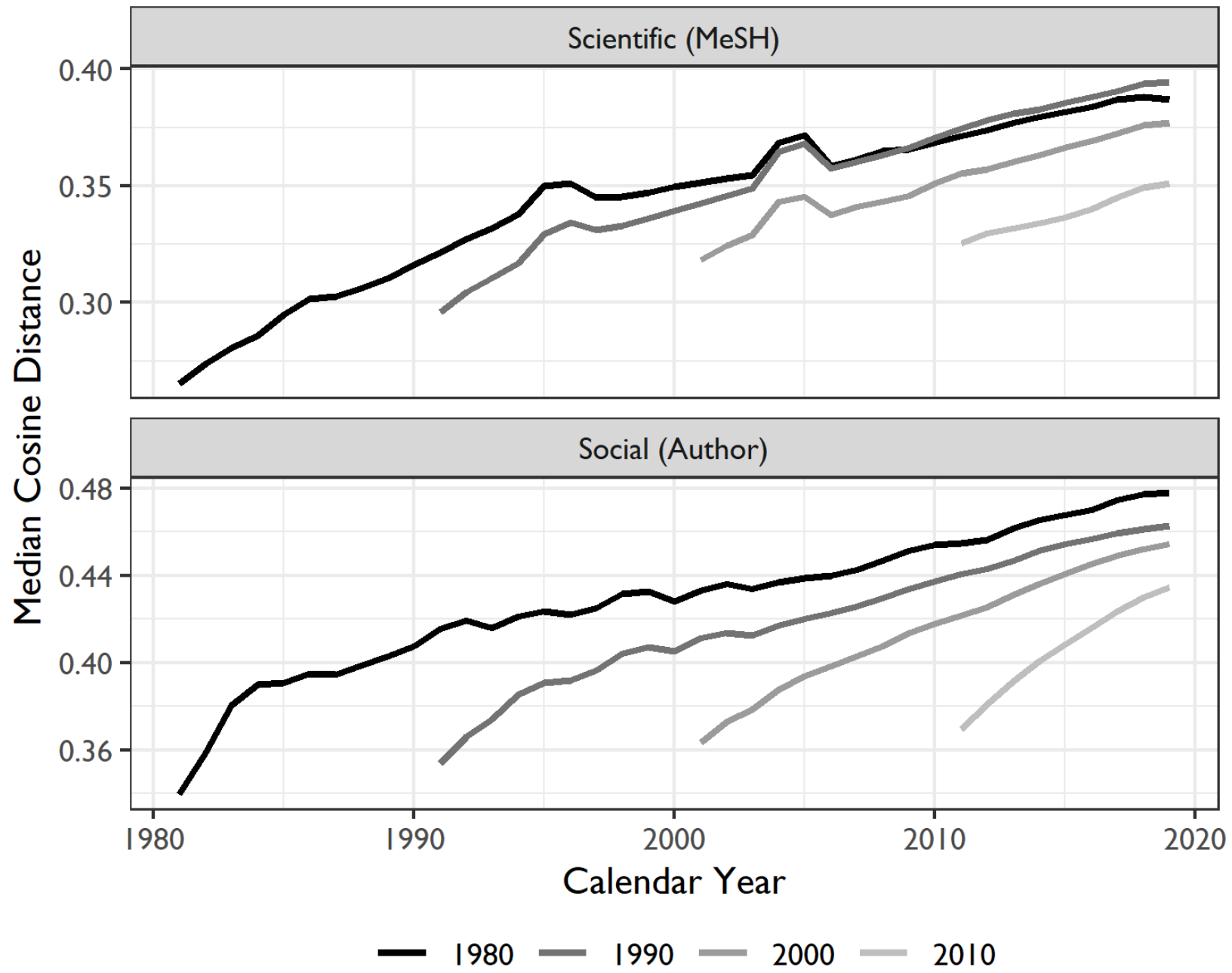

**Fig. S2.2: Temporal pattern of diffusion from highly cited articles (Top 5%) published in 1980, 1990, 2000, 2010**. Median cosine distances for each year (t) are computed based on a two-year rolling (t and t-1) window.

Fig. S2.2 shows the temporal evolution of diffusion metrics from scientific and social space. As the universe of biomedical entities and scientists expands, distances between focal papers and citing papers tend to increase in both scientific and social spaces by 2019. Then, the pattern, especially from the scientific space, indicates that our 100-dimensional representational spaces may allocate publications in some years (e.g., 2004 and 2005) in relatively distant locations within a trained manifold in the training process. Together, these suggest a necessity to consider the calendar year effect when a research article was published for the following analysis.

## S3. Delineating Biomedical Subfields

### *S3.1 Dataset for Seed Articles and PMRA*

Science is a social enterprise: like any other intellectual product, biomedical science attains its significance when others recognize and build upon it[11–13]. Hence, we seek to understand the dynamics of diffusion and shifting attention beyond individual publications at the subfield level. We utilize the 'Similar Article' (or 'Related Articles') function provided by PubMed[14], powered by the Pubmed Related Article algorithm (PMRA)[14], which uses words in the abstracts, titles, and MeSH terms to capture a set of intellectually neighboring articles from a given seed article. This approach has been employed to study the repercussions of scientific scandals on careers[15], shifts in research focus among scientists responding to NIH funding changes[15,16], and the negative impact of winning prizes for recipient competitors[17].

We apply this method to seed articles curated by Azoulay and their colleagues[18], which consist of research papers from U.S. elite life scientists published between 1970 and 2002 (inclusive). The authors deemed scientists elite if they satisfied one or more of the following criteria: they were (i) highly funded, (ii) highly cited, (iii) top patents, (iv) members of the National Academy of Sciences, (v) the National Academy of Medicine, or (vi) early career prize winners (i.e., NIH MERIT awardees, Howard Hughes Medical Investigators). To estimate the effects of the premature death of elite biomedical researchers, the study first identified 3,076 seed articles authored by 452 researchers who died prematurely. These 452 researchers represent a subset of a larger pool of 12,935 star scientists, with a median (and mean) age at death of 61. Within this group, 229 passed away following a protracted illness, while 185 died suddenly and unexpectedly (e.g., in a car crash). Forty percent of these stars held an M.D. rather than a Ph.D.; 90 percent were male, and each received an average of $16.6 million in NIH grants and published 138 papers, garnering 8,341 citations over their careers.

The study then performed 'coarsened exact matching' to identify a control group of publications from elite scientists who did not experience premature death, taking into account factors such as 1) publication years, 2) team sizes, 3) the ages of the elite scientists, and 4) long-run citation impact. This process allows a subfield to be matched to several subfields associated with the early death of eminent scientists, leading to overlaps. Consequently, 4,180 subfields were matched to more than one subfield with the premature death of associated superstar scientists, leading to duplicates when counted individually. This results in 34,218 unique pairs of subfield strata and seed articles for subfield identification, for 28,504 unique seed articles spanning subfields. We repurpose this dataset to investigate a different question from the original study in our work.

Consistent with the original approach, we assume each seed article represents distinct subfields in biomedicine, but we extended the period to the end of 2019 as their subfield panel data stopped in 2006. We identify 1,941,680 unique publications (including the seed articles) spanning 28,504 unique subfields. By the end of 2019, the mean and median size of subfields is 122.52 and 102, respectively, with a standard deviation of 91.5. Table S3.1 below shows the subfield sizes at the 1st, 10th, 25th (1Q), 50th (median), 75th (3Q), 90th, and 99th percentiles at the end of 2019. Table S3.2 shows the distribution of citations garnered by 28,504 subfields by the end of 2019.

**Table S3.1:** Distribution of Subfield Size in 2019.

| Percentile | 1th | 10th | 25th | 50th | 75th | 90th | 99th |
|---|---|---|---|---|---|---|---|
| Subfield Size in 2019 | 40 | 68 | 89 | 102 | 125 | 184 | 484 |

**Table S3.2:** Distribution of Cumulative Citations per Subfield at the End of 2019.

| Percentile | 1th | 10th | 25th | 50th | 75th | 90th | 99th |
|---|---|---|---|---|---|---|---|
| Cumulative # of citations by the end of 2019 | 1,015 | 2,319 | 3,423 | 5,097 | 7,506 | 11,065 | 24,312 |

We extract research articles that have cited any 1,941,680 publications from the PKG 2020 citation database, which returns 11,421,194 publications and 86,804,637 paper-to-paper citations. Not all publications are associated with MeSH or Author IDs from PKG. We identify 10,894,779 publications that PKG assigns author IDs, constituting 84,389,548 citations from social space and 10,454,104 publications associated with MeSH terms linked through 82,228,828 citations.

*S3.2 Alternative Identification of Subfields*

While PMRA underpins the PubMed interface, serving as a crucial tool for researchers to locate information related to focal research papers and study the operation of biomedical science[15–18], our analysis relies on this specific method of subfield identification. To ensure the robustness of our results, we have undertaken the following steps: 1) Using the same set of 28,504 seed article PMIDs and the scientific embedding space trained on MeSH terms, we redefined subfields by selecting the top $N$ most similar articles to the seed articles based on cosine similarity within the scientific space, where '$N$' corresponds to the original subfield size as determined by PMRA. For the 82 out of 28,504 seed articles without assigned MeSH terms, we substituted each with a PMRA-identified similar article that ranked in the top 10 in similarity and had the smallest difference in publication years; 2) As an additional robustness check, we doubled the size of each subfield to assess the sensitivity of our results to changes in subfield sizes identified by PMRA.

With these two alternatively defined systems of subfields, we recalculated subfield-level variables consistent with our original analysis. We applied the same approach using three thresholds (i.e., 0.5%, 0.25%, 0.1%) to identify the sudden declines from two alternative subfields of the same and doubled sizes defined within our scientific embedding space. The following tables report a pattern of results similar to the main findings.

**Table S3.3:** Estimates with Alternative Subfields using the Same Size of the Originals.

| **Dependent Variable** | ***Substantial Decline of Citation*** | | |
|---|---|---|---|
| | **Model 1 (0.5%)** | **Model 2 (0.25%)** | **Model 3 (0.1%)** |
| ***Knowledge Diffusion*** | | | |
| Scientific Space | -0.250***<br>[-0.312, -0.188]<br>($p < 0.001$) | -0.289***<br>[-0.364, -0.213]<br>($p < 0.001$) | -0.358***<br>[-0.445, -0.270]<br>($p < 0.001$) |
| Social Space | -0.098**<br>[-0.167, -0.03]<br>($p = 0.005$) | -0.113*<br>[-0.205, -0.02]<br>($p = 0.017$) | -0.150*<br>[-0.293, -0.006]<br>($p = 0.041$) |
| **Log-Likelihood** | -33,501.0 | -19,308.5 | -8,780.9 |
| **Total Observations** | 1,324,948 | 1,385,980 | 1,425,731 |

**Note:** *Knowledge Diffusion*, *Subfield Growth Pattern*, *and Citation Dynamics* are all one-year lagged. The diffusion indices are standardized within field ages and calendar years across 28,504 subfields. The 95% confidence intervals inside brackets are computed based on the standard errors clustered at strata ID and calendar years. Coefficients for field ages are omitted. * $p < .05$; ** $p < .01$; *** $p < .001$ (two-tailed)

**Table S3.4:** Estimates with Controls for Alternative Subfields using the Same Size.

| **Dependent Variable** | | ***Substantial Decline of Citation*** | |
|---|---|---|---|
| | **Model 1 (0.5%)** | **Model 2 (0.25%)** | **Model 3 (0.1%)** |
| ***Knowledge Diffusion*** | | | |
| Scientific Space | -0.258***<br>[-0.321, -0.194]<br>($p$ < 0.001) | -0.276***<br>[-0.356, -0.196]<br>($p$ < 0.001) | -0.339***<br>[-0.471, -0.208]<br>($p$ < 0.001) |
| Social Space | -0.121**<br>[-0.204, -0.038]<br>($p$ = 0.004) | -0.157**<br>[-0.263, -0.052]<br>($p$ = 0.003) | -0.201*<br>[-0.360, -0.043]<br>($p$ = 0.013) |
| ***Subfield Growth*** | | | |
| Cum. Subfield Size (logged) | 0.686***<br>[0.539, 0.833]<br>($p$ < 0.001) | 0.781***<br>[0.561, 1.000]<br>($p$ < 0.001) | 0.716***<br>[0.388, 1.043]<br>($p$ < 0.001) |
| 2-year Subfield Growth | -0.237***<br>[-0.256, -0.217]<br>($p$ < 0.001) | -0.244***<br>[-0.269, -0.218]<br>($p$ < 0.001) | -0.258***<br>[-0.291, -0.225]<br>($p$ < 0.001) |
| ***Citation Dynamics*** | | | |
| Cum. Citations (logged) | -2.113***<br>[-2.426, -1.800]<br>($p$ < 0.001) | -2.291***<br>[-2.648, -1.933]<br>($p$ < 0.001) | -2.443***<br>[-2.863, -2.024]<br>($p$ < 0.001) |
| 2-year Citations (logged) | 2.152***<br>[1.881, 2.422]<br>($p$ < 0.001) | 2.278***<br>[1.960, 2.595]<br>($p$ < 0.001) | 2.440***<br>[2.042, 2.838]<br>($p$ < 0.001) |
| Gini Coef. of Cum. Citation | 0.262<br>[-0.986, 1.510]<br>($p$ = 0.680) | 0.341<br>[-1.466, 2.148]<br>($p$ = 0.711) | -0.261<br>[-3.222, 2.700]<br>($p$ = 0.863) |
| Gini Coef. of 2-year Citation | -0.598<br>[-1.688, 0.492]<br>($p$ = 0.282) | -0.748<br>[-2.457, 0.961]<br>($p$ = 0.391) | -1.046<br>[-3.976, 1.883]<br>($p$ = 0.484) |
| ***Other Controls*** | | | |
| Retraction Notice Published | -0.494<br>[-1.062, 0.074]<br>($p$ = 0.088) | -0.262<br>[-1.22, 0.696]<br>($p$ = 0.592) | 0.299<br>[-0.873, 1.471]<br>($p$ = 0.617) |
| After Death | 0.191*<br>[0.034, 0.348]<br>($p$ = 0.017) | 0.420***<br>[0.221, 0.619]<br>($p$ < 0.001) | 0.413<br>[-0.005, 0.831]<br>($p$ = 0.053) |
| After Death * Superstar Death | 0.017<br>[-0.143, 0.177]<br>($p$ = 0.836) | 0.027<br>[-0.169, 0.222]<br>($p$ = 0.788) | -0.094<br>[-0.428, 0.240]<br>($p$ = 0.581) |
| ***Fixed Effects*** | | | |
| Calendar Year | Y | Y | Y |
| Strata ID | Y | Y | Y |
| **Log-Likelihood** | -27,714.8 | -15,542.6 | -6,594.8 |
| **Total Observations** | 1,324,948 | 1,385,980 | 1,425,731 |

**Note:** *Knowledge Diffusion*, *Subfield Growth Pattern*, *and Citation Dynamics* are all one-year lagged. The diffusion indices are standardized within field ages and calendar years across 28,504 subfields. The 95% confidence intervals inside brackets are computed based on the standard errors clustered at strata ID and calendar years. Coefficients for field ages are omitted. * $p < .05$; ** $p < .01$; *** $p < .001$ (two-tailed)

**Table S3.5:** Estimates with Subfields from Scientific Space with Doubled Size of the Originals.

| **Dependent Variable** | ***Substantial Decline of Citation*** | | |
|---|---|---|---|
| | **Model 1 (0.5%)** | **Model 2 (0.25%)** | **Model 3 (0.1%)** |
| ***Knowledge Diffusion*** | | | |
| Scientific Space | -0.268***<br>[-0.359, -0.177]<br>($p < 0.001$) | -0.344***<br>[-0.442, -0.247]<br>($p < 0.001$) | -0.395***<br>[-0.536, -0.255]<br>($p < 0.001$) |
| Social Space | -0.144***<br>[-0.215, -0.073]<br>($p < 0.001$) | -0.141**<br>[-0.247, -0.034]<br>($p = 0.009$) | -0.196*<br>[-0.360, -0.031]<br>($p = 0.02$) |
| **Log-Likelihood** | -34,063.1 | -19,601.3 | -8,827.1 |
| **Total Observations** | 1,389,103 | 1,446,414 | 1,484,148 |

**Note:** *Knowledge Diffusion*, *Subfield Growth Pattern*, *and Citation Dynamics* are all one-year lagged. The diffusion indices are standardized within field ages and calendar years across 28,504 subfields. The 95% confidence intervals inside brackets are computed based on the standard errors clustered at strata ID and calendar years. Coefficients for field ages are omitted. * $p < .05$; ** $p < .01$; *** $p < .001$ (two-tailed)

**Table S3.6:** Estimates with Controls for Alternative Subfields using the Double Size.

| Dependent Variable | *Substantial Decline of Citation* | | |
|---|---|---|---|
| | **Model 1 (0.5%)** | **Model 2 (0.25%)** | **Model 3 (0.1%)** |
| ***Knowledge Diffusion*** | | | |
| Scientific Space | -0.288***<br>[-0.364, -0.212]<br>($p < 0.001$) | -0.345***<br>[-0.432, -0.258]<br>($p < 0.001$) | -0.355***<br>[-0.523, -0.186]<br>($p < 0.001$) |
| Social Space | -0.178***<br>[-0.258, -0.097]<br>($p < 0.001$) | -0.208***<br>[-0.323, -0.093]<br>($p < 0.001$) | -0.265**<br>[-0.439, -0.091]<br>($p = 0.003$) |
| ***Subfield Growth*** | | | |
| Cum. Subfield Size (logged) | 0.603***<br>[0.422, 0.784]<br>($p < 0.001$) | 0.679***<br>[0.447, 0.91]<br>($p < 0.001$) | 0.883***<br>[0.487, 1.279]<br>($p < 0.001$) |
| 2-year Subfield Growth | -0.307***<br>[-0.332, -0.282]<br>($p < 0.001$) | -0.320***<br>[-0.347, -0.293]<br>($p < 0.001$) | -0.352***<br>[-0.401, -0.304]<br>($p < 0.001$) |
| ***Citation Dynamics*** | | | |
| Cum. Citations (logged) | -2.542***<br>[-2.880, -2.204]<br>($p < 0.001$) | -2.738***<br>[-3.126, -2.349]<br>($p < 0.001$) | -3.204***<br>[-3.583, -2.825]<br>($p < 0.001$) |
| 2-year Citations (logged) | 2.532***<br>[2.241, 2.824]<br>($p < 0.001$) | 2.678***<br>[2.334, 3.022]<br>($p < 0.001$) | 3.029***<br>[2.649, 3.409]<br>($p < 0.001$) |
| Gini Coef. of Cum. Citation | 1.989**<br>[0.671, 3.306]<br>($p = 0.003$) | 2.409*<br>[0.52, 4.298]<br>($p = 0.012$) | 2.447<br>[-1.217, 6.111]<br>($p = 0.191$) |
| Gini Coef. of 2-year Citation | -1.634*<br>[-2.951, -0.317]<br>($p = 0.015$) | -1.896<br>[-3.932, 0.141]<br>($p = 0.068$) | -2.831<br>[-6.915, 1.252]<br>($p = 0.174$) |
| ***Other Controls*** | | | |
| Retraction Notice Published | -0.193<br>[-0.606, 0.221]<br>($p = 0.362$) | -0.102<br>[-0.76, 0.556]<br>($p = 0.762$) | 0.290<br>[-0.776, 1.356]<br>($p = 0.594$) |
| After Death | 0.200<br>[0.006, 0.393]<br>($p = 0.043$) | 0.188<br>[-0.101, 0.477]<br>($p = 0.202$) | 0.306<br>[-0.185, 0.797]<br>($p = 0.222$) |
| After Death * Superstar Death | -0.104<br>[-0.252, 0.043]<br>($p = 0.166$) | -0.115<br>[-0.316, 0.087]<br>($p = 0.264$) | -0.152<br>[-0.45, 0.145]<br>($p = 0.315$) |
| ***Fixed Effects*** | | | |
| Calendar Year | Y | Y | Y |
| Strata ID | Y | Y | Y |
| **Log-Likelihood** | -27,851.1 | -15,571.3 | -6,521.8 |
| **Total Observations** | 1,389,103 | 1,446,414 | 1,484,148 |

**Note:** *Knowledge Diffusion*, *Subfield Growth Pattern*, *and Citation Dynamics* are all one-year lagged. The diffusion indices are standardized within field ages and calendar years across 28,504 subfields. The 95% confidence intervals inside brackets are computed based on the standard errors clustered at strata ID and calendar years. Coefficients for field ages are omitted. * $p < .05$; ** $p < .01$; *** $p < .001$ (two-tailed)

*S3.3 Robustness Check With Mutually Exclusive Subfields*

Using seed articles from the Azoulay team, the PMRA-based subfield identification method allows a paper to be included in more than one subfield. The Azoulay team explored the extent of shared articles between pairs of PMRA-delineated subfields, focusing their analysis on 21,661 subfield pairs where a deceased superstar was last author on both associated source articles (see Figure C6 at aeaweb.org/content/file?id=10303). This suggests that although overlaps occur, most articles are associated with a single subfield delineated by PMRA. We conducted our investigation, noting that 1) the Azoulay team's analysis focused only on subfields associated with scientists who died prematurely, and 2) we extended the analysis window up to 2019. Our analysis shows that 64.1% of papers are predominantly associated with a single subfield.

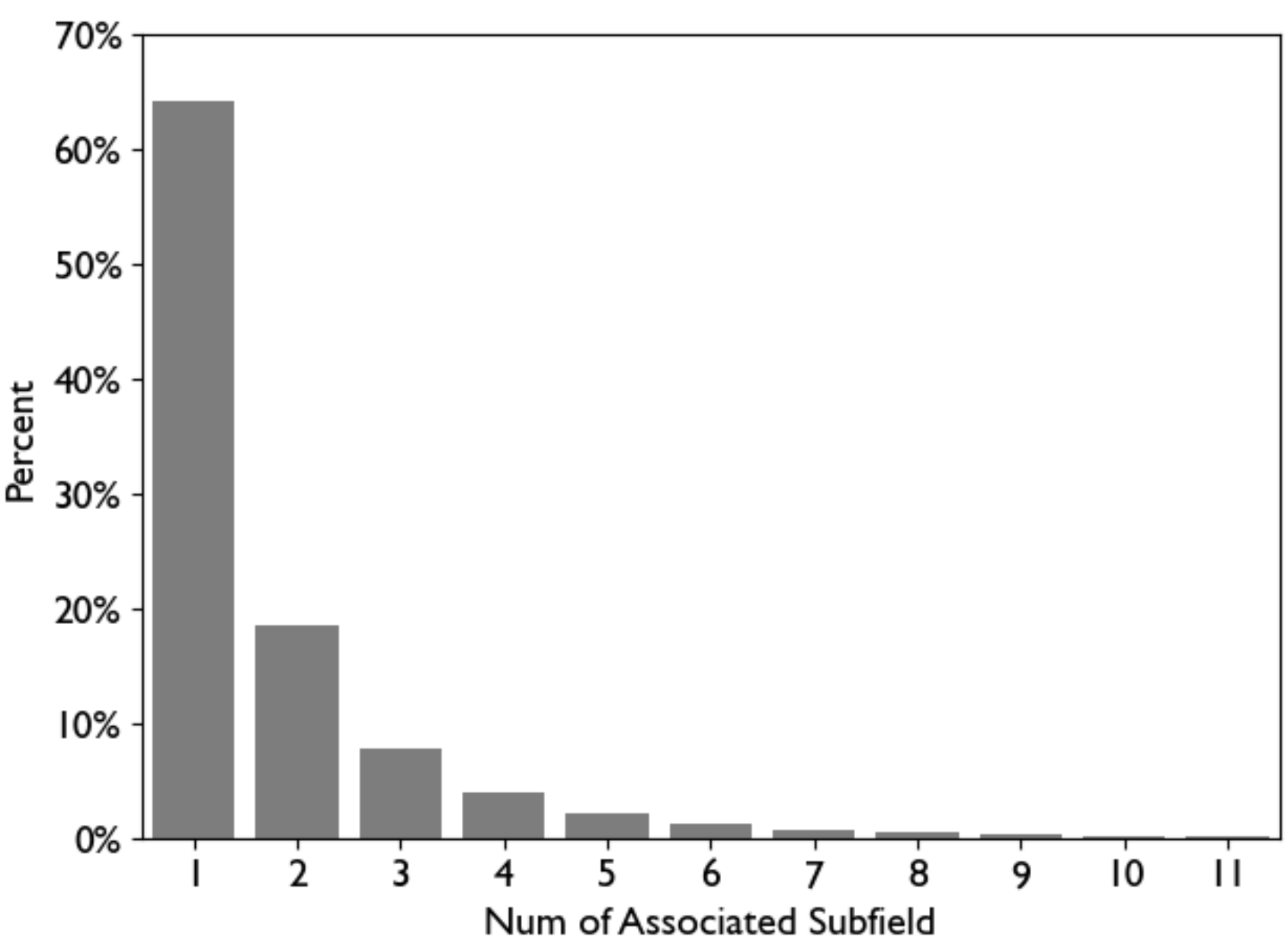

**Fig. S3.1: Proportion of Papers by Number of Subfield Associated.**

We conducted additional analyses by reassigning overlapping papers to a single seed article, ensuring each field is mutually exclusive. We did this by leveraging distances within our scientific space. Specifically, for papers associated with more than one seed article, we compute cosine similarities between each paper and its seed articles and select the seed article with the highest similarity. Following this reassignment, we recomputed all subfield-level variables, including diffusion metrics, outcomes, and controls, in the same manner as in our original analyses. Results confirmed the robustness of our findings, as detailed below.

**Table S3.7:** Model Estimates after Reassignment Using the Bottom 0.50% Cutoff.

| **Dependent Variable** | *Substantial Decline of Citation* | | | | | |
|---|---|---|---|---|---|---|
| | **Model 1** | **Model 2** | **Model 3** | **Model 4** | **Model 5** | **Model 6** |
| ***Knowledge Diffusion*** | | | | | | |
| Scientific Space | -0.204***<br>[-0.267, -0.140]<br>$p < 0.001$ | -0.232***<br>[-0.293, -0.170]<br>$p < 0.001$ | -0.128***<br>[-0.171, -0.086]<br>$p < 0.001$ | -0.129***<br>[-0.171, -0.088]<br>$p < 0.001$ | -0.182***<br>[-0.233, -0.131]<br>$p < 0.001$ | -0.218***<br>[-0.277, -0.160]<br>$p < 0.001$ |
| Social Space | -0.131***<br>[-0.185, -0.077]<br>$p < 0.001$ | -0.189***<br>[-0.249, -0.129]<br>$p < 0.001$ | -0.150***<br>[-0.211, -0.089]<br>$p < 0.001$ | -0.148***<br>[-0.208, -0.088]<br>$p < 0.001$ | -0.163***<br>[-0.229, -0.096]<br>$p < 0.001$ | -0.161***<br>[-0.227, -0.095]<br>$p < 0.001$ |
| ***Subfield Growth*** | | | | | | |
| Cum. Subfield Size (logged) | | 0.309***<br>[0.249, 0.369]<br>$p < 0.001$ | 0.563***<br>[0.466, 0.661]<br>$p < 0.001$ | 0.572***<br>[0.471, 0.673]<br>$p < 0.001$ | 0.151*<br>[0.014, 0.287]<br>$p = 0.031$ | 0.049<br>[-0.071, 0.170]<br>$p = 0.421$ |
| 2-year Subfield Growth | | -0.094***<br>[-0.104, -0.084]<br>$p < 0.001$ | -0.143***<br>[-0.157, -0.129]<br>$p < 0.001$ | -0.143***<br>[-0.157, -0.130]<br>$p < 0.001$ | -0.136***<br>[-0.150, -0.122]<br>$p < 0.001$ | -0.142***<br>[-0.158, -0.127]<br>$p < 0.001$ |
| ***Citation Dynamics*** | | | | | | |
| Cum. Citations (logged) | | | -1.895***<br>[-2.154, -1.637]<br>$p < 0.001$ | -1.906***<br>[-2.168, -1.645]<br>$p < 0.001$ | -1.650***<br>[-1.955, -1.345]<br>$p < 0.001$ | -1.673***<br>[-1.957, -1.389]<br>$p < 0.001$ |
| 2-year Citations (logged) | | | 1.909***<br>[1.655, 2.163]<br>$p < 0.001$ | 1.914***<br>[1.658, 2.169]<br>$p < 0.001$ | 2.210***<br>[1.941, 2.479]<br>$p < 0.001$ | 2.426***<br>[2.154, 2.698]<br>$p < 0.001$ |
| Gini Coef. of Cum. Citation | | | 0.432<br>[-0.225, 1.089]<br>$p = 0.198$ | 0.461<br>[-0.196, 1.119]<br>$p = 0.169$ | 0.025<br>[-0.741, 0.791]<br>$p = 0.949$ | 0.180<br>[-0.597, 0.957]<br>$p = 0.65$ |
| Gini Coef. of 2-year Citation | | | -0.511<br>[-1.131, 0.110]<br>$p = 0.107$ | -0.516<br>[-1.139, 0.108]<br>$p = 0.105$ | -1.420***<br>[-2.096, -0.745]<br>$p < 0.001$ | -1.593***<br>[-2.230, -0.956]<br>$p < 0.001$ |
| ***Other Controls*** | | | | | | |
| Retraction Notice Published | | | -0.237<br>[-0.649, 0.175]<br>$p = 0.26$ | -0.077<br>[-0.531, 0.376]<br>$p = 0.738$ | 0.01<br>[-0.54, 0.56]<br>$p = 0.972$ | -0.237<br>[-0.649, 0.175]<br>$p = 0.26$ |
| After Death | | | 0.080<br>[-0.121, 0.28]<br>$p = 0.437$ | 0.121<br>[-0.031, 0.274]<br>$p = 0.118$ | 0.2*<br>[0.032, 0.369]<br>$p = 0.02$ | 0.08<br>[-0.121, 0.28]<br>$p = 0.437$ |
| After Death * Superstar Death | | | -0.149*<br>[-0.283, -0.015]<br>$p = 0.029$ | -0.002<br>[-0.168, 0.165]<br>$p = 0.984$ | -0.51**<br>[-0.895, -0.125]<br>$p = 0.009$ | -0.149*<br>[-0.283, -0.015]<br>$p = 0.029$ |
| ***Fixed Effects*** | | | | | | |
| Calendar Year | N | N | N | N | Y | Y |
| Strata ID | N | N | N | N | Y | N |
| Star ID | N | N | N | N | N | Y |
| **Log-Likelihood** | -30,045.0 | -29,187.9 | -27,699.2 | -27,694.0 | -25,118.1 | -22,709.7 |
| **Total Observations** | 1,252,242 | | | | | |

**Note:** *Knowledge Diffusion*, *Subfield Growth Pattern*, *and Citation Dynamics* are all one-year lagged. The diffusion indices are standardized within field ages and calendar years across 28,504 subfields. The 95% confidence intervals inside brackets are computed based on the standard errors clustered at star's ID and calendar years. Coefficients for field ages are omitted. * $p < .05$; ** $p < .01$; *** $p < .001$ (two-tailed)

**Table S3.8:** Model Estimates after Reassignment Using the Bottom 0.25% Cutoff.

| **Dependent Variable** | *Substantial Decline of Citation* | | | | | |
|---|---|---|---|---|---|---|
| | **Model 1** | **Model 2** | **Model 3** | **Model 4** | **Model 5** | **Model 6** |
| ***Knowledge Diffusion*** | | | | | | |
| Scientific Space | -0.195***<br>[-0.273, -0.116]<br>$p < 0.001$ | -0.217***<br>[-0.295, -0.14]<br>$p < 0.001$ | -0.102***<br>[-0.159, -0.045]<br>$p < 0.001$ | -0.103***<br>[-0.159, -0.046]<br>$p < 0.001$ | -0.142***<br>[-0.211, -0.074]<br>$p < 0.001$ | -0.171***<br>[-0.244, -0.098]<br>$p < 0.001$ |
| Social Space | -0.168***<br>[-0.231, -0.105]<br>$p < 0.001$ | -0.236***<br>[-0.308, -0.164]<br>$p < 0.001$ | -0.195***<br>[-0.268, -0.121]<br>$p < 0.001$ | -0.193***<br>[-0.266, -0.12]<br>$p < 0.001$ | -0.196***<br>[-0.276, -0.117]<br>$p < 0.001$ | -0.193***<br>[-0.272, -0.114]<br>$p < 0.001$ |
| ***Subfield Growth*** | | | | | | |
| Cum. Subfield Size (logged) | | 0.357***<br>[0.288, 0.427]<br>$p < 0.001$ | 0.696***<br>[0.581, 0.811]<br>$p < 0.001$ | 0.701***<br>[0.583, 0.819]<br>$p < 0.001$ | 0.289***<br>[0.116, 0.461]<br>$p = 0.001$ | 0.155<br>[-0.023, 0.333]<br>$p = 0.089$ |
| 2-year Subfield Growth | | -0.097***<br>[-0.109, -0.086]<br>$p < 0.001$ | -0.15***<br>[-0.166, -0.135]<br>$p < 0.001$ | -0.151***<br>[-0.166, -0.136]<br>$p < 0.001$ | -0.144***<br>[-0.16, -0.129]<br>$p < 0.001$ | -0.151***<br>[-0.167, -0.135]<br>$p < 0.001$ |
| ***Citation Dynamics*** | | | | | | |
| Cum. Citations (logged) | | | -2.090***<br>[-2.347, -1.833]<br>$p < 0.001$ | -2.097***<br>[-2.353, -1.84]<br>$p < 0.001$ | -1.850***<br>[-2.153, -1.547]<br>$p < 0.001$ | -1.921***<br>[-2.241, -1.601]<br>$p < 0.001$ |
| 2-year Citations (logged) | | | 2.046***<br>[1.795, 2.297]<br>$p < 0.001$ | 2.049***<br>[1.798, 2.300]<br>$p < 0.001$ | 2.357***<br>[2.079, 2.635]<br>$p < 0.001$ | 2.658***<br>[2.360, 2.956]<br>$p < 0.001$ |
| Gini Coef. of Cum. Citation | | | 0.665<br>[-0.110, 1.439]<br>$p = 0.092$ | 0.681<br>[-0.090, 1.452]<br>$p = 0.083$ | 0.082<br>[-0.790, 0.955]<br>$p = 0.854$ | 0.503<br>[-0.419, 1.424]<br>$p = 0.285$ |
| Gini Coef. of 2-year Citation | | | -0.759*<br>[-1.48, -0.038]<br>$p = 0.039$ | -0.763*<br>[-1.485, -0.04]<br>$p = 0.039$ | -1.650***<br>[-2.449, -0.852]<br>$p < 0.001$ | -2.077***<br>[-2.848, -1.305]<br>$p < 0.001$ |
| ***Other Controls*** | | | | | | |
| Retraction Notice Published | | | -0.193<br>[-0.725, 0.339]<br>$p = 0.476$ | 0.079<br>[-0.451, 0.609]<br>$p = 0.771$ | -0.007<br>[-0.7, 0.685]<br>$p = 0.983$ | -0.193<br>[-0.725, 0.339]<br>$p = 0.476$ |
| After Death | | | 0.050<br>[-0.218, 0.318]<br>$p = 0.713$ | 0.049<br>[-0.224, 0.321]<br>$p = 0.727$ | 0.064<br>[-0.200, 0.328]<br>$p = 0.634$ | 0.050<br>[-0.218, 0.318]<br>$p = 0.713$ |
| After Death * Superstar Death | | | -0.090<br>[-0.294, 0.115]<br>$p = 0.391$ | 0.115<br>[-0.129, 0.359]<br>$p = 0.356$ | -0.151<br>[-0.652, 0.349]<br>$p = 0.554$ | -0.09<br>[-0.294, 0.115]<br>$p = 0.391$ |
| ***Fixed Effects*** | | | | | | |
| Calendar Year | N | N | N | N | Y | Y |
| Strata ID | N | N | N | N | Y | N |
| Star ID | N | N | N | N | N | Y |
| **Log-Likelihood** | -16,969.8 | -16,495.6 | -15,626.5 | -15,625.3 | -13,682.5 | -12,015.4 |
| **Total Observations** | 1,304,469 | | | | | |

**Note:** *Knowledge Diffusion*, *Subfield Growth Pattern*, *and Citation Dynamics* are all one-year lagged. The diffusion indices are standardized within field ages and calendar years across 28,504 subfields. The 95% confidence intervals inside brackets are computed based on the standard errors clustered at star's ID and calendar years. Coefficients for field ages are omitted. * $p < .05$; ** $p < .01$; *** $p < .001$ (two-tailed)

**Table S3.9:** Model Estimates after Reassignment Using the Bottom 0.10% Cutoff.

| **Dependent Variable** | *Substantial Decline of Citation* | | | | | |
|---|---|---|---|---|---|---|
| | **Model 1** | **Model 2** | **Model 3** | **Model 4** | **Model 5** | **Model 6** |
| ***Knowledge Diffusion*** | | | | | | |
| Scientific Space | -0.243***<br>[-0.373, -0.113]<br>$p < 0.001$ | -0.265***<br>[-0.400, -0.129]<br>$p < 0.001$ | -0.139*<br>[-0.251, -0.028]<br>$p = 0.014$ | -0.140*<br>[-0.251, -0.028]<br>$p = 0.014$ | -0.161*<br>[-0.310, -0.013]<br>$p = 0.033$ | -0.229***<br>[-0.367, -0.091]<br>$p = 0.001$ |
| Social Space | -0.198***<br>[-0.298, -0.098]<br>$p < 0.001$ | -0.271***<br>[-0.384, -0.158]<br>$p < 0.001$ | -0.229***<br>[-0.344, -0.114]<br>$p < 0.001$ | -0.226***<br>[-0.341, -0.111]<br>$p < 0.001$ | -0.238***<br>[-0.368, -0.109]<br>$p < 0.001$ | -0.186*<br>[-0.331, -0.040]<br>$p = 0.012$ |
| ***Subfield Growth*** | | | | | | |
| Cum. Subfield Size (logged) | | 0.385***<br>[0.294, 0.475]<br>$p < 0.001$ | 0.778***<br>[0.613, 0.943]<br>$p < 0.001$ | 0.786***<br>[0.618, 0.954]<br>$p < 0.001$ | 0.326*<br>[0.052, 0.601]<br>$p = 0.02$ | 0.225<br>[-0.099, 0.55]<br>$p = 0.173$ |
| 2-year Subfield Growth | | -0.097***<br>[-0.114, -0.079]<br>$p < 0.001$ | -0.151***<br>[-0.170, -0.132]<br>$p < 0.001$ | -0.151***<br>[-0.171, -0.132]<br>$p < 0.001$ | -0.146***<br>[-0.167, -0.125]<br>$p < 0.001$ | -0.154***<br>[-0.175, -0.133]<br>$p < 0.001$ |
| ***Citation Dynamics*** | | | | | | |
| Cum. Citations (logged) | | | -2.178***<br>[-2.514, -1.843]<br>$p < 0.001$ | -2.188***<br>[-2.525, -1.85]<br>$p < 0.001$ | -1.940***<br>[-2.350, -1.529]<br>$p < 0.001$ | -2.125***<br>[-2.594, -1.656]<br>$p < 0.001$ |
| 2-year Citations (logged) | | | 2.101***<br>[1.760, 2.443]<br>$p < 0.001$ | 2.109***<br>[1.765, 2.453]<br>$p < 0.001$ | 2.435***<br>[2.068, 2.803]<br>$p < 0.001$ | 2.849***<br>[2.444, 3.253]<br>$p < 0.001$ |
| Gini Coef. of Cum. Citation | | | 0.411<br>[-0.928, 1.751]<br>$p = 0.547$ | 0.425<br>[-0.911, 1.762]<br>$p = 0.533$ | -0.053<br>[-1.657, 1.552]<br>$p = 0.949$ | 0.594<br>[-0.958, 2.147]<br>$p = 0.453$ |
| Gini Coef. of 2-year Citation | | | -0.822<br>[-2.041, 0.396]<br>$p = 0.186$ | -0.838<br>[-2.061, 0.386]<br>$p = 0.18$ | -1.742**<br>[-3.049, -0.435]<br>$p = 0.009$ | -2.515***<br>[-3.858, -1.172]<br>$p < 0.001$ |
| ***Other Controls*** | | | | | | |
| Retraction Notice Published | | | | -1.018<br>[-2.166, 0.13]<br>$p = 0.082$ | -0.614<br>[-1.912, 0.684]<br>$p = 0.354$ | -0.551<br>[-2.007, 0.905]<br>$p = 0.458$ |
| After Death | | | | 0.172<br>[-0.231, 0.575]<br>$p = 0.402$ | 0.289<br>[-0.123, 0.701]<br>$p = 0.17$ | 0.232<br>[-0.168, 0.633]<br>$p = 0.255$ |
| After Death * Superstar Death | | | | -0.124<br>[-0.402, 0.154]<br>$p = 0.381$ | -0.024<br>[-0.395, 0.346]<br>$p = 0.898$ | 0.297<br>[-0.948, 1.542]<br>$p = 0.64$ |
| ***Fixed Effects*** | | | | | | |
| Calendar Year | N | N | N | N | Y | Y |
| Strata ID | N | N | N | N | Y | N |
| Star ID | N | N | N | N | N | Y |
| **Log-Likelihood** | -7,706.9 | -7,511.0 | -7,141.6 | -7,138.5 | -5,858.2 | -4,901.2 |
| **Total Observations** | 1,337,848 | | | | | |

**Note:** *Knowledge Diffusion*, *Subfield Growth Pattern*, *and Citation Dynamics* are all one-year lagged. The diffusion indices are standardized within field ages and calendar years across 28,504 subfields. The 95% confidence intervals inside brackets are computed based on the standard errors clustered at star's ID and calendar years. Coefficients for field ages are omitted. * $p < .05$; ** $p < .01$; *** $p < .001$ (two-tailed)

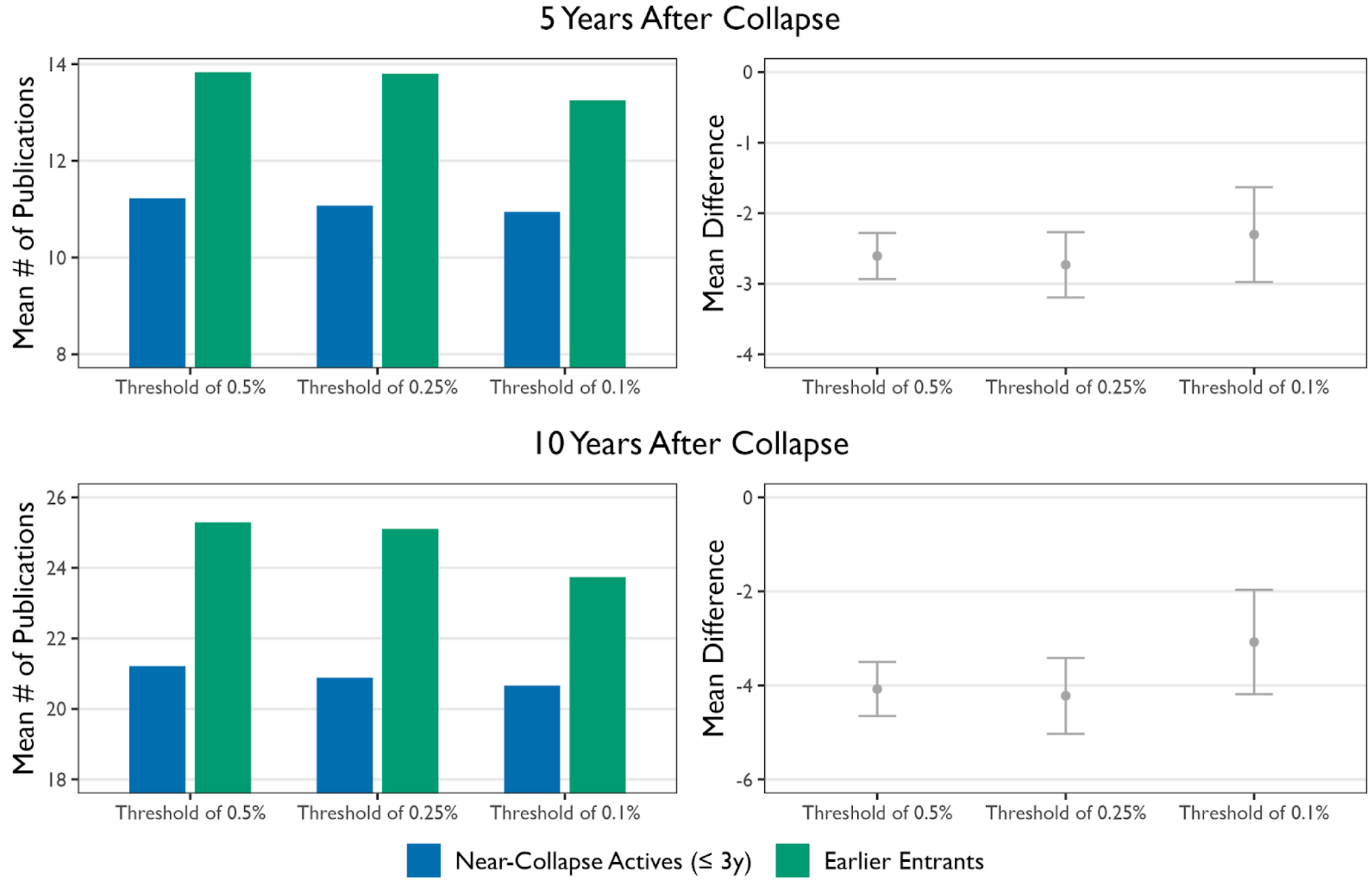

**Fig. S3.2: Comparing author productivity in collapsed subfields 5 and 10 years post-collapse after reassignment.** Error bars represent 95% confidence intervals for the mean differences in average publication numbers. Comparisons are drawn between authors who entered the field early and those active near the collapse, based on paired *t*-tests.

**Table S3.10:** Pairwise *t*-test comparing average productivity differences between near-collapse active scientists (≤2 years before collapse) and early entrants after reassignment**.**

| | **Threshold** | **Estimate** | **t** | ***p*-value (*d.f.*)** | **95% C.I.** |
|---|---|---|---|---|---|
| **5 Years** | 0.5% | -2.607 | -15.60 | < 0.001 (3,236) | [-2.280, -2.935] |
| | 0.25% | -2.731 | -11.57 | < 0.001 (1,673) | [-2.269, -3.195] |
| | 0.1% | -2.303 | -6.72 | < 0.001 (655) | [-1.631, -2.976] |
| **10 Years** | 0.5% | -4.077 | -13.87 | < 0.001 (2,972) | [-3.500,-4.653] |
| | 0.25% | -4.222 | -10.25 | < 0.001 (1,547) | [-3.415,-5.030] |
| | 0.1% | -3.078 | -5.45 | < 0.001 (603) | [-1.968,-4.187] |

**Note**: Subfields that collapsed after 2015 were excluded from the 5-year productivity comparison. Likewise, for the 10-year productivity, only subfields that collapsed on or before 2011 were included, considering the observation windows.

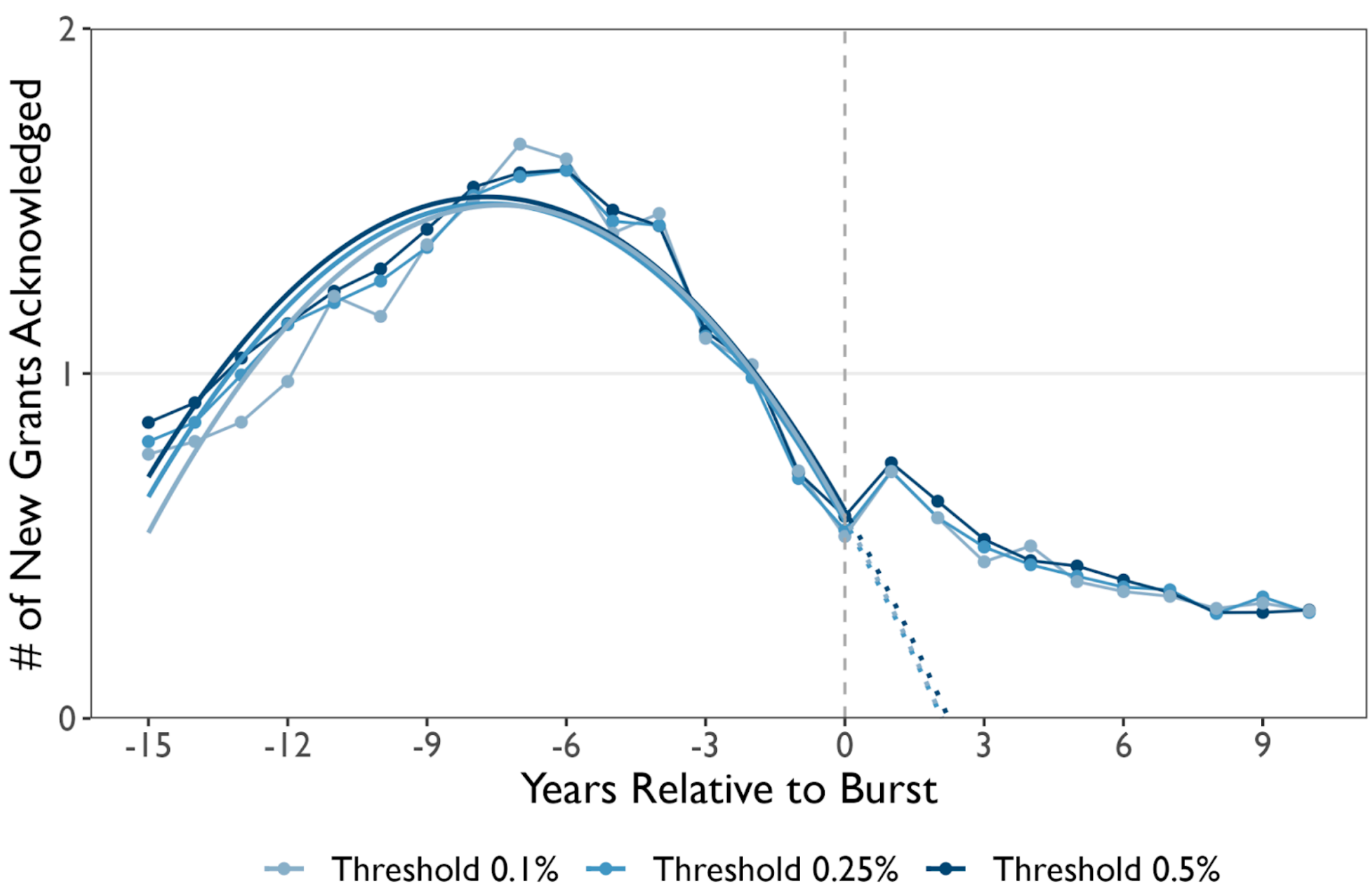

**Fig. S3.3: Average number of new grants acknowledged in collapsed subfields by years relative to burst after reassignment.** The quadratic fit is applied to data from years -15 to 0 relative to the burst year, with dotted lines representing extrapolations starting from year 0 onwards.

**Table S3.11:** Proportion of subfields with newly acknowledged grants after collapse, and the Mean, 1st Quantile, Median, and 3rd Quartile of the number of new grants post-collapse after reassignment.

| Threshold | % of Subfields with New Grants Acknowledged After Collapse | Mean | Q1 | Median | Q3 |
|---|---|---|---|---|---|
| 0.5% | 68.45% | 6.0 | 0 | 2 | 7 |
| 0.25% | 68.79% | 5.9 | 0 | 2 | 7 |
| 0.1% | 68.68% | 5.8 | 0 | 2 | 6 |

**Note:** Subfields that collapsed after 2015 were excluded from the 5-year productivity comparison. For the 10-year productivity analysis, only subfields that collapsed on or before 2011 were included, in consideration of the observation windows.

**S4. Post-Hoc Analyses**

To gain further insights into the phenomenon of scientific bubbles, we complement our analysis with augmented data, including subfield-level characteristics initially provided by the Azoulay team and information extracted from the NIH's iCite system[19,20].

*S4.1 Stars' Importance*

In exploring the mechanisms associated with the likelihood of scientific bubbles bursting, as suggested by the Stem Cell Cardiac regeneration case, we examine the relationship between 'Star Importance to the Subfield' and the likelihood of subfield collapses. To do so, we draw on the replication data provided by the Azoulay team, defining 'Star Importance to the Subfield' as the fraction of papers authored by superstar scientists within the subfield (variable name: 'imprtnc'). Utilizing logistic regression, we assess whether this measure can predict the likelihood of collapses versus uncollapsed subfields. As detailed in S3, while the original dataset contains 28,504 unique seed articles, it yields 34,218 pairs of subfield strata and seed articles for subfield identification. Thus, we applied clustered standard errors at the Strata IDs. Table S4.1 shows a significant association between the stars' importance and the likelihood of collapse; Fig. S4.1 visualizes the estimation reported in Table 4.1, which suggests the association between concentration of scientific capital and the probability of a subfield collapsing.

**Table S4.1:** Star's Importance to the Subfield and Collapse.

| **Dependent Variable** | ***Collapsed versus Not Collapsed*** | | |
|---|---|---|---|
| **Threshold** | **0.5%** | **0.25%** | **0.1%** |
| (Intercept) | -1.862***<br>[-1.922, -1.802]<br>($p < 0.001$) | -2.636***<br>[-2.706, -2.565]<br>($p < 0.001$) | -3.627***<br>[-3.725, -3.528]<br>($p < 0.001$) |
| Star's Importance to the Subfield | 1.222***<br>[1.001, 1.443]<br>($p < 0.001$) | 1.299***<br>[1.038, 1.560]<br>($p < 0.001$) | 1.320***<br>[0.961, 1.680]<br>($p < 0.001$) |
| **Log-Likelihood** | -14,899.2 | -9,588.9 | -4,814.2 |
| **Total Observations (# of Unique Seed Article-Strata Pairs)** | | 34,218 | |

**Note:** The 95% confidence intervals and p-values are based on the standard errors clustered at strata ID.
* $p < .05$; ** $p < .01$; *** $p < .001$ (two-tailed)

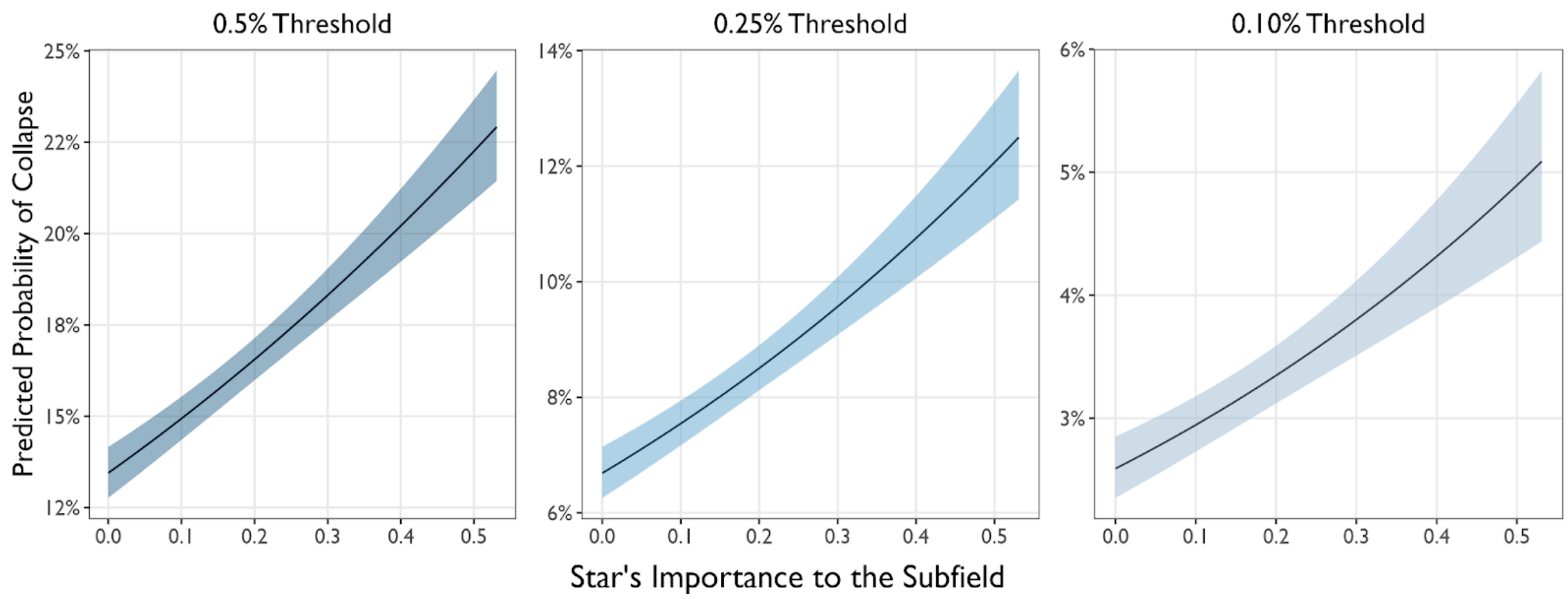

**Fig. S4.1: Predicted Probability of Collapse by the Star's Importance to the Subfield, Based on the Estimates in Table S4.1.** The range of Star's Importance to the Subfield extends up to 3 standard deviations from the distribution of the variable. The mean and standard deviation of the variable, Star's Importance to the Subfield, are 0.151 and 0.127, respectively. The bands represent the 95% confidence intervals, calculated based on the standard errors reported in Table S4.1.

*S4.2 Subfield Funding Accounted by Star's Collaborators*

We additionally examine the relationship between the 'Fraction of Subfield Funding Accounted by Star's Collaborators' and the likelihood of subfield collapses, using the same approach presented in S4.1 for 'Star's Importance.' This analysis assesses the association between the 'fraction of subfield funding accounted for by collaborators' (variable name: 'frac_collabs_field_nih_fndg') and collapses. Table S4.2 demonstrates a significant association between the concentration of funding among star scientists' collaborators and the collapse. Fig. S4.2 visualizes the estimation reported in Table 4.2, further suggesting an association between the concentration of scientific capital and the probability of a subfield collapsing.

**Table S4.2:** Fraction of Subfield Funding Accounted by Star's Collaborators.

| **Dependent Variable** | ***Collapsed versus Not Collapsed*** | | |
|---|---|---|---|
| **Threshold** | **0.5%** | **0.25%** | **0.1%** |
| (Intercept) | -1.734***<br>[-1.791, -1.676]<br>(*p* < 0.001) | -2.484***<br>[-2.556, -2.412]<br>(*p* < 0.001) | -3.481***<br>[-3.581, -3.381]<br>(*p* < 0.001) |
| Fraction of Subfield Funding Accounted by Star's Collaborators | 0.224***<br>[0.110, 0.337]<br>(*p* < 0.001) | 0.194*<br>[0.037, 0.350]<br>(*p* = 0.015) | 0.231*<br>[0.004, 0.458]<br>(*p* = 0.046) |
| **Log-Likelihood** | -14,951.6 | -9,626.2 | -4,830.1 |
| **Total Observations (# of Unique Seed Article-Strata Pairs)** | 34,218 | | |

**Note:** The 95% confidence intervals and p-values are based on the standard errors clustered at strata ID.
* $p < .05$; ** $p < .01$; *** $p < .001$ (two-tailed)

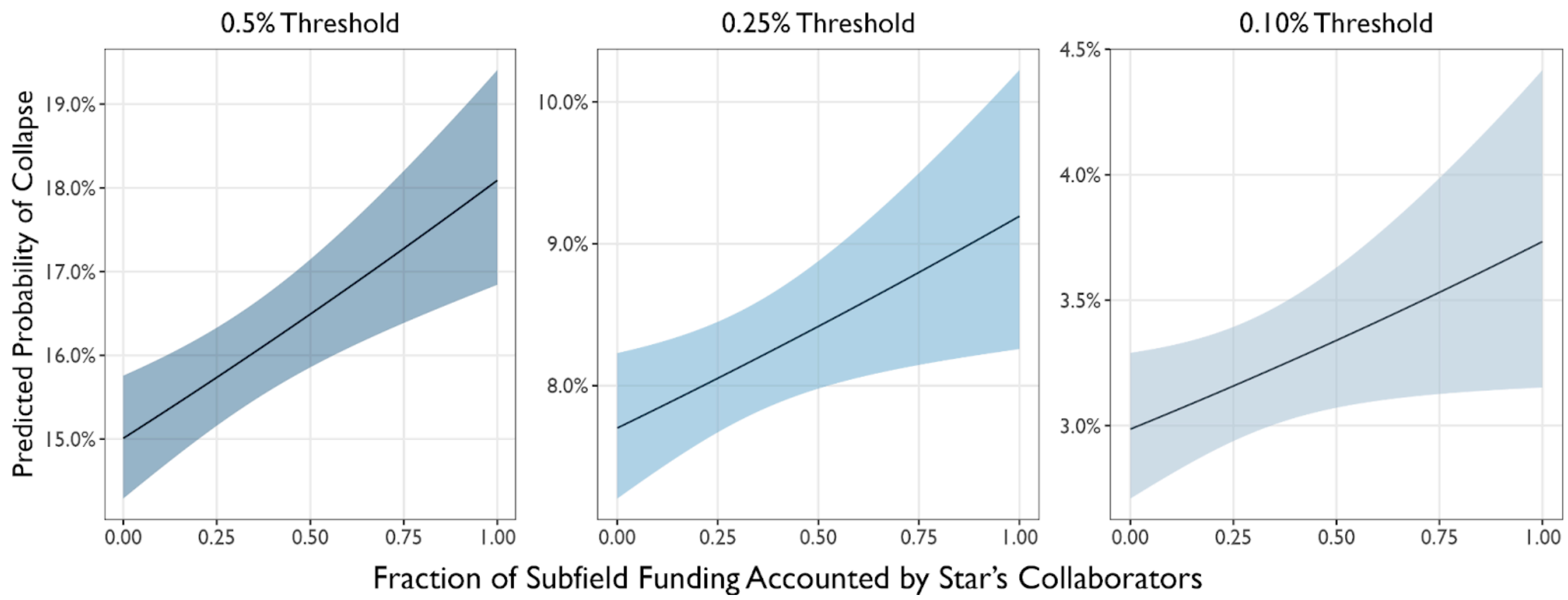

**Fig. S4.2: Predicted Probability of Collapse by Fraction of Subfield Funding Accounted by Star's Collaborators, Based on the Estimates in Table S4.2.** The mean and standard deviation of the variable, Fraction of Subfield Funding Accounted by Star's Collaborators, are 0.283 and 0.293, respectively. The bands represent the 95% confidence intervals, calculated based on the standard errors reported in Table S4.2.

*S4.3 Approximate Potential for Clinical Translation*

We further augmented our dataset with the Approximate Potential for Clinical Translation (APT) modules[19] from PubMed's iCite system. This module evaluates the likelihood of a paper's research being applied and cited in subsequent clinical studies. We extracted APT values for each publication in our dataset from the iCite bulk data. These values were then aggregated at the subfield level by calculating the average APT for articles cited at least once up to a specific calendar year. We selected the year when collapsed subfields experienced collapses and matched this time to the calendar year (and consequently the field age) of subfields that did not collapse within the strata initially established by the Azoulay team. The subsequent logistic regression, using average APT values aggregated by calendar year and field age as predictors, indicates that a higher potential for clinical translation is negatively associated with the likelihood of a subfield experiencing a drastic collapse, as Table S4.3 and Fig S4.3 report. This suggests that bubbles may be associated with limited clinical relevance, and conversely, greater clinical relevance appears to prevent such collapses.

**Table S4.3:** Approximate Potential to Clinical Translation.

| **Dependent Variable** | ***Collapsed versus Not Collapsed*** | | |
|---|---|---|---|
| **Threshold** | **0.5%** | **0.25%** | **0.1%** |
| (Intercept) | 0.124***<br>[0.09, 0.158]<br>($p < 0.001$ | 0.142***<br>[0.097, 0.188]<br>($p < 0.001$) | 0.136***<br>[0.059, 0.213]<br>($p < 0.001$) |
| Approximate Potential to Clinical Translation | -0.509***<br>[-0.672, -0.347]<br>($p < 0.001$) | -0.650***<br>[-0.875, -0.424]<br>($p < 0.001$) | -0.658***<br>[-1.047, -0.268]<br>($p < 0.001$) |
| **Log-Likelihood** | -6,073.3 | -3,348.3 | -1,365.8 |
| **Total Observations (Matched Year-Field Age-Strata)** | 8,773 | 4,840 | 1,974 |

**Note:** The 95% confidence intervals and p-values are based on the standard errors clustered at strata ID-Field age pair. * $p < .05$; ** $p < .01$; *** $p < .001$ (two-tailed)

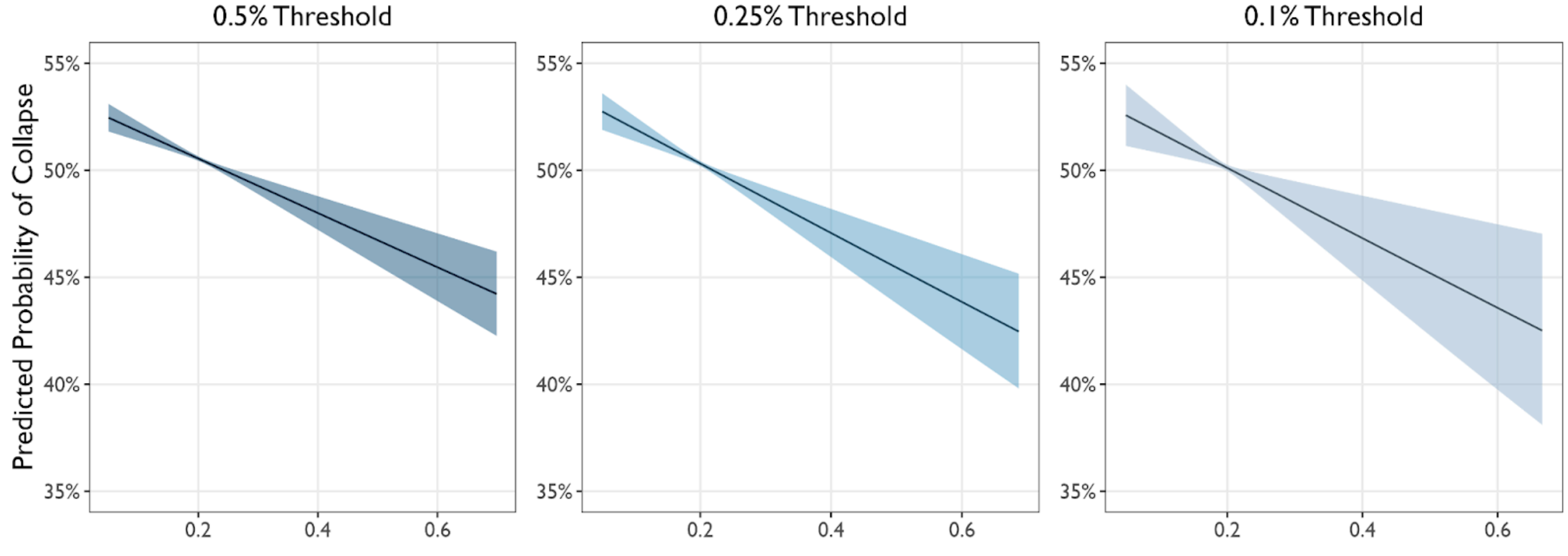

**Fig. S4.3: Predicted Probability of Collapse by the Average Approximate Potential to Clinical Translation of Cited Papers within the Subfield, Based on the Estimates in Table S4.3.** The range of Average Approximate Potential to Clinical Translation of Cited Papers within the Subfield extends up to 3 standard deviations from the distribution. The bands represent the 95% confidence intervals, calculated based on the standard errors reported in Table S4.3.

*S4.4 Comparing Actual vs. Expected Citations in Collapsed and Uncollapsed Subfields*

To further support the concept of bubble as 'inflated' attention and our operationalization of it, we investigate how the citation counts that a subfield garners before collapse deviate from expected citations, comparing these discrepancies between collapsed and uncollapsed subfields.

We utilized the NIH's iCite Influence Module, which provides an annual 'expected citation count' for each MEDLINE-indexed paper. This system offers a benchmark for the number of citations a typical (median) MEDLINE-indexed paper, identified from the co-citation network and published in the same year, would receive[20]. After extracting the expected annual citation count for all publications in our dataset, we summed these expected citation counts for each subfield across each calendar year for the articles published up to those years. This approach enabled us to compute the difference between the actual citations a subfield garnered and the expected count. We then focused on the one and two years before the collapse of subfields. We matched these years with subfields that did not collapse within the same strata (i.e., same publication year, similar long-term citation counts, similar star-scientist age, distinct intellectual location), analogous to S4.3.

Using logistic regressions, with average deviation from expected citation counts aggregated by calendar year and strata to capture the degree of "inflation" as a predictor, the analysis presented below in Table S4.4 and Fig. 4.4 suggests that a greater degree of deviation from expected citation counts is positively associated with the likelihood of a subfield experiencing a collapse in the subsequent year. This pattern represents a positive indication of the attention bubble that may subsequently burst. When combined with results from the main analysis, this indicates that positive deviations, or the indication of bubbles, are inversely correlated with diffusion.

**Table S4.4:** Difference between Actual and Expected Citations Before Collapse (2y).

| **Dependent Variable** | ***Collapsed versus Not Collapsed*** | | |
|---|---|---|---|
| **Threshold** | **0.5%** | **0.25%** | **0.1%** |
| (Intercept) | -0.238***<br>[-0.273, -0.203]<br>($p < 0.001$) | -0.270***<br>[-0.317, -0.223]<br>($p < 0.001$) | -0.273***<br>[-0.346, -0.199]<br>($p < 0.001$) |
| Difference between Actual and Expected Citations Before Collapse (2y) | 0.002***<br>[0.002, 0.002]<br>($p < 0.001$) | 0.002***<br>[0.002, 0.003]<br>($p < 0.001$) | 0.002***<br>[0.002, 0.003]<br>($p < 0.001$) |
| **Log-Likelihood** | -6,016.7 | -3,314.1 | -1,352.8 |
| **Total Observations (Matched Year-Field Age-Strata)** | 8,773 | 4,840 | 1,974 |

**Note:** The 95% confidence intervals and p-values are based on the standard errors clustered at strata ID-Field age pair. * $p < .05$; ** $p < .01$; *** $p < .001$ (two-tailed)

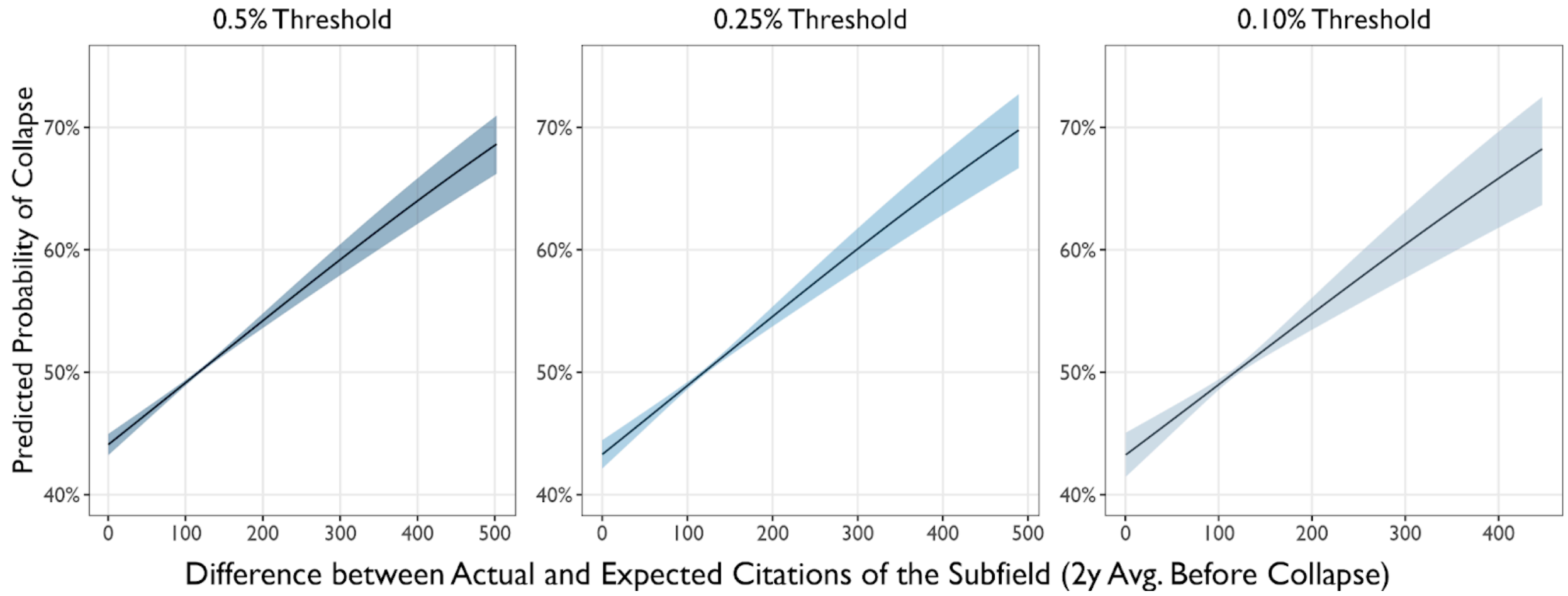

**Fig. S4.4: Predicted Probability of Collapse by the Difference Between Actual and Expected Citation, Based on the Estimates in Table S4.4.** The range of Difference Between Actual and Expected Citation, Based extends up to 3 standard deviations from the distribution. The bands represent the 95% confidence intervals, calculated based on the standard errors reported in Table S4.4.

## References (for Supplementary Information)